%% file: ms.tex
\documentclass[10pt,preprint]{aastex}
\usepackage{amsmath}
\usepackage{amssymb}

\newcommand{\A}[1]{{\bf #1}}
\newcommand{\pd}[2]{\frac{\partial #1}{\partial #2}}
\newcommand{\HALF}{\frac{1}{2}}
\newcommand{\DS}{\displaystyle}
\newcommand{\td}{\tilde{\delta}_{\textrm I}}

\slugcomment{Submitted to the \apj}
\shorttitle{Dynamics of Radiative Shocks}
\shortauthors{Mignone}
\received{2004 October 19}

\begin{document}

\title{THE DYNAMICS OF RADIATIVE SHOCK WAVES: LINEAR AND NONLINEAR EVOLUTION}
\author{A. Mignone\altaffilmark{1,2,3}}

\altaffiltext{1}{INAF Osservatorio Astronomico di Torino, Strada dell'Osservatorio 20, 10025 Pino Torinese, Italy}
\altaffiltext{2}{Department of Astronomy \& Astrophysics,
   The University of Chicago,
   Chicago, IL 60637}
\altaffiltext{3}{Center for Astrophysical Thermonuclear Flashes,
   The University of Chicago,
   Chicago, IL 60637}

\begin{abstract}
 The stability properties of one-dimensional radiative shocks with a power-law
 cooling function of the form $\Lambda \propto \rho^2T^\alpha$ are the main
 subject of this work.
 The linear analysis originally presented by Chevalier \& 
 Imamura, is thoroughfully reviewed for several values of the cooling 
 index $\alpha$ and higher overtone modes. 
 Consistently with previous results, it is shown that
 the spectrum of the linear operator consists in a series of modes with 
 increasing oscillation frequency.
 For each mode a critical value of the cooling index, $\alpha_{\textrm c}$, 
 can be defined so that modes with $\alpha < \alpha_{\textrm c}$
 are unstable, while modes with $\alpha > \alpha_{\textrm c}$ are stable. 
 The perturbative analysis is complemented by several numerical 
 simulations to follow the time-dependent evolution of 
 the system for different values of $\alpha$.
 Particular attention is given to the comparison between
 numerical and analytical results (during the early phases
 of the evolution) and to the role played by different boundary
 conditions. It is shown that an appropriate treatment of the lower 
 boundary yields results that closely follow the predicted 
 linear behavior. During the nonlinear regime, the 
 shock oscillations saturate at a finite amplitude and tend
 to a quasi-periodic cycle.
 The modes of oscillations during this phase do not necessarily 
 coincide with those predicted by linear theory, but may be accounted for 
 by mode-mode coupling.

% For the first time, evidence is provided in favour of nonlinear
% modes being supported by mode-mode coupling.

\end{abstract}


%%%%%%%%%%%%%%%%%%%%%%%%%%%%%%%%%%%%%%%%%%%%%%%%%%%%%%%%%%%%%%%%%%%%%%
\section{INTRODUCTION}
%
%
%
%
%
%%%%%%%%%%%%%%%%%%%%%%%%%%%%%%%%%%%%%%%%%%%%%%%%%%%%%%%%%%%%%%%%%%%%%%

Radiative shock waves are believed to play a key
role in a variety of different astrophysical environments, including
magnetic cataclysmic variables  \citep{Wu00,Crop90},
jets from young stellar objects \citep{Hart94},
magnetospheric accretion in T-Tauri stars \citep{CG98},  
colliding stellar winds \citep{Stevens92, AOB04} and supernova 
remnants \citep{KC97, Blondin98, WF98}.

Most of the earlier theoretical investigation has been  
motivated by the dynamics of accreting shocks in magnetic
cataclysmic variables.
In these systems, a strongly magnetized white dwarf 
($10^6\lesssim B \lesssim 10^8$ Gauss) accretes matter directly 
from a late-type star without the formation of a disk.
Instead, the mass transfer process is magnetically channeled and
matter accumulates through a stand-off shock on a small fraction of 
the stellar surface.
At temperatures of $10^8\div 10^9$ K, the shock-heated plasma
radiates its energy via optically thin bremsstrahlung becoming a 
powerful source of X-rays.

It has been shown that, under certain circumstances, the 
postshock flow is subject to a global thermal instability 
(more precisely, an overstability) caused by rapid variations of 
the cooling time scale with the shock speed. 
The instability drives the shock front to oscillate with
respect to its stationary position, causing variations in 
the amount of radiation emitted from the postshock region.
The instability mechanism has been invoked in the past to explain
the optical quasi-periodic-oscillations (QPO) observed in 
AM Her-type systems \citep{Larsson92, Middleditch97, Wu00}.
Similarly, a relevant issue arises in questioning the validity of 
steady shock models with shock velocities $v_{\textrm s} \gtrsim 130$ km s$^{-1}$,
routinely used in interpreting emission line 
observations from interstellar shocks \citep{IGF87a,IGF87b,GEC88,SBD03}.

The nature of the instability was first studied analytically
by \citet[CI hereafter]{CI82}, who presented a linear stability 
analysis of planar radiative shocks with volumetric cooling rate 
$\Lambda\propto\rho^2 T^\alpha$.
CI showed that the shock has multiple modes of oscillation, and 
the stability of a particular mode depends on the value of the
cooling index $\alpha$.
In general, higher power dependences on the temperature were
shown to inhibit the growth of instability.
Thus the fundamental mode was shown to become unstable for 
$\alpha\lesssim 0.4$, while the first and second harmonics are 
de-stabilized when $\alpha\lesssim 0.8$. Higher order harmonics
were not considered by CI.

Perturbative studies of one-temperature flows with a power-law
cooling function were afterwards considered by several authors.
\citet{Ima85} presented a linear stability study of radiative 
shocks where the cooling function had a weaker dependence on
density, i.e. $\Lambda\propto\rho T^\alpha$.
\citet{Bert86} examined the structure of spherical radiative 
shocks and also considered the effects of nonradial perturbations.
He showed that modes that are stable to radial perturbations 
may become unstable for small transverse wavenumbers. 
For sufficiently large wavenumbers, however, 
all modes are eventually stabilized.

Effects due to cyclotron emission were considered in 
\citet{Ima91}, \citet{Wu92}, \citet{Wu96}.
Noise driven models were proposed by \citet{Wood92}.
Effects of gravity were studied by \citet{HC92}, while 
magnetic field effects were considered by 
\citet{Ed89a}, \citet{Ed89b}, \citet{TD93}, and by \citet{HP98}
in the two-dimensional case.
\citet{DS94} considered radiative shocks with a mass loss
term.

Recently, \citet{YM01} investigated the stability properties 
of shock-compressed gas slabs by introducing a cold layer 
of finite thickness. 
They considered both symmetric and antisymmetric modes and
reported the existence of quasi-oscillatory modes, in addition
to the overstable ones.

Stability properties of radiative shocks with unequal ion and
electron temperatures were investigated by \citet{Ima96}, while 
flows with multiple cooling functions (i.e. bremsstrahlung and 
cyclotron) were examined both in the single and two-temperature
regimes by \citet{Sax97}, \citet{Sax98}, \citet{Sax99}, 
\citet{SaxWu99}, \citet{SaxWu01} and \citet{Sax02}. Additional
references can be found in the review by \citet{Wu00}.

Investigation of the full time-dependent problem has also received 
extensive attention over the past two decades and several 
numerical simulations have been carried along.
The oscillatory instability was discovered in the first place 
numerically by \citet[LCSa hereafter]{LCS81} and 
\citet[LCSb hereafter]{LCS82}, who investigated spherically 
symmetric accretion onto magnetized white dwarfs.
They showed that flows with volumetric cooling rates $\Lambda\sim\rho^2T^\alpha$ 
become unstable when $\alpha \lesssim 1.6$, a limit subsequently
revised to $\alpha \lesssim 0.6$ in \citet{LCS83}. 
In the case of bremsstrahlung cooling ($\alpha = 1/2$), the shock position 
was shown to undergo periodic oscillations over the surface of the
white dwarf. 
\citet[IWD hereafter]{Ima84} and \citet{Ima85} used a Lagrangian code 
to investigate the stability of radiative shocks with a power-law cooling 
function $\Lambda\sim \rho^2T^\alpha$. Effects of gravity, thermal
conduction and unequal electron and ion temperatures were also included.
In IWD, the critical values of $\alpha$ (above which oscillations are damped) 
were shown to lie somewhere in the range $1/3\lesssim\alpha\lesssim 1/2$
and $1/2\lesssim \alpha\lesssim 0.6$ for the fundamental and first-overtone, 
respectively.
\citet{Wolff89} considered both one and two-temperature calculations;
the one-temperature calculations showed that the fundamental mode
is unstable for $\alpha = 0.33$, first and second harmonic are unstable
when $\alpha = 0.65$, while the system is stable at $\alpha = 1$.
The one-dimensional calculations of \citet[SB hereafter]{SB95} 
showed that for flows incident into a wall, large amplitude 
oscillations are damped when $\alpha\gtrsim 0.5$.
SB also considered perturbed steady-state models, showing that power-law 
cooling functions with $\alpha \lesssim 0.75$ produced shock oscillations 
which saturated at a finite amplitude.
The results of SB were recently extended by \citet[SBD hereafter]{SBD03} to a more
realistic cooling function.

Most results of the previous numerical investigations,
however, bear no clear relation with the predicted linear behavior,
and a direct comparison with perturbative studies has proven 
to be only partially successful.
Although the salient features of thermally unstable shocks have been 
commonly reproduced, the modes of instabilities can not always be  
identified with the linear ones, with the exception of one or two modes.
Besides, controversies exist on the value of the critical $\alpha$ 
above which the system should become stable.
These apparent inconsistencies may be partially due to the fact that 
most calculations do not include a stationary solution in their 
initial condition, which makes the problem inherently nonlinear 
since the very beginning.
Moreover, a substantial disagreement exists between Eulerian and
Lagrangian calculations and on the numerical treatment of the lower
boundary condition which plays a crucial role in the dynamics of the 
post shock flow.

Here I wish to present some new and detailed calculations in 
the attempt to settle part of these controversies. The results of this work
will also serve as a basis for future extensions, where effects of
magnetic fields and more realistic cooling functions will be
considered.

The paper is organized as follows. In \S\ref{sec:eqns}
the problem is defined and the relevant equations are introduced.
In \S\ref{sec:linear} the stability properties 
of 1-D planar radiative shocks are reviewed in more detail, 
while in \S\ref{sec:time_dependent} numerical simulations
are presented with particular emphasis to the comparison 
with linear theory and to the choice of boundary conditions.
A new time-dependent treatment of the lower boundary is introduced 
and the details of implementation are given in Appendix \ref{app:cvbc}.

%%%%%%%%%%%%%%%%%%%%%%%%%%%%%%%%%%%%%%%%%%%%%%%%%%%%%%%%%%%%%%%%%%%%%%
\section{STATEMENT OF THE PROBLEM AND EQUATIONS} \label{sec:eqns}
%
%
%
%
%
%%%%%%%%%%%%%%%%%%%%%%%%%%%%%%%%%%%%%%%%%%%%%%%%%%%%%%%%%%%%%%%%%%%%%%

Consider a one-dimensional supersonic flow with uniform density
$\rho_{\textrm{in}}$ and velocity $v_{\textrm{in}}$, initially propagating in 
the negative x-direction, i.e., $v_{\textrm{in}} = - |v_{\textrm{in}}|$.
The flow is brought to rest by the presence of
a rigid wall located at $x=0$, and a shock wave forms
at some finite distance $x_{\textrm{s}}$ from the wall.
Through the shock, the bulk kinetic energy of the 
incoming gas is converted into thermal motion and the flow
decelerates to subsonic velocities.
In the postshock region, the thermal energy of the
accreting gas is radiated away by cooling processes.

In steady-state, the dynamical time scale is equal to the cooling time
scale, so a fluid element travels through the postshock region
and cools exactly to zero temperature by the time it reaches the wall.

In several areas of interest and to make the problem 
more tractable, radiative losses are treated in the 
optically thin regime and the cooling rates are normally 
specified as functions of the temperature, density and 
relative abundances. 
In this work it is assumed that the volumetric cooling rates
can be described by a single power-law (in temperature) function
\begin{equation}\label{eq:lambda}
 \Lambda(\rho,p) =  a \rho^2\left(\frac{p}{\rho}\right)^\alpha \,,
\end{equation}
where $a$ is a physical constant depending on the
particular cooling process, $\alpha$ is the cooling
index and $\rho$ and $p$ are, respectively, the density and
pressure of the gas.
 
The problem of a supersonic flow into a wall is, of course, a simplified 
abstraction of a more complex and specific physical setting. 
Effects due to thermal conduction, magnetic fields and 
multi-dimensional effects are neglected in this paper
and will be considered in future works.
With these assumptions, the problem can be
described by the Euler equations for a one-temperature 
flow in planar geometry:
\begin{equation}\label{eq:continuity}
 \pd{\rho}{t} + \rho\pd{v}{x} + v\pd{\rho}{x} = 0 \,,
\end{equation}
\begin{equation}\label{eq:momentum}
 \pd{v}{t} + v\pd{v}{x} + \frac{1}{\rho}\pd{p}{x} = 0 \,,
\end{equation}
\begin{equation}\label{eq:energy}
  \pd{p}{t} + v\pd{p}{x} + \gamma p \pd{v}{x} = 
  - (\gamma-1){\cal C} \rho^2\left(\frac{p}{\rho}\right)^\alpha \,,
\end{equation}
where $v$ is the fluid velocity, ${\cal C}$ is a constant and
an ideal equation of state with constant specific heat ratio 
$\gamma$ has been assumed.
Equations (\ref{eq:continuity}) through (\ref{eq:energy}) are
put in a dimensionless form by expressing density and velocity 
in units of their inflow values, i.e., $\rho_{\textrm{in}}$
and $|v_{\textrm{in}}|$.
With this choice, the flow variables immediately ahead of the
shock become $\rho=1$, $v=-1$ and $p = 1/(\gamma\mathcal{M}^2)$,
with $\mathcal{M}$ being the upstream Mach number.

The length scale of the problem enters explicitly through the 
constant ${\cal C}$ in the energy equation (\ref{eq:energy}):
\begin{equation}\label{eq:normalization}
 {\cal C} = L_{\textrm c} a \rho_{\textrm{in}}^{1 - 2\alpha}|v_{\textrm{in}}|^{2\alpha - 3} \,,
\end{equation}
where $L_{\textrm c}$ sets the reference length scale
and $a$ has already been introduced in equation (\ref{eq:lambda}).
In the following, lengths will be conveniently normalized to
the stationary thickness of the cooling region, so that
the equilibrium position of the shock is $x=1$.
The explicit expression for ${\cal C}$ is given in
\S\ref{sec:perturb}.

Relations between quantities ahead and behind the shock
follow from the Rankine-Hugoniot jump conditions:
\begin{equation}\label{eq:rankine_1}
 -v_{\textrm s}   = \frac{1}{\rho_{\textrm s}} = \frac{\gamma - 1}{\gamma + 1} +
          \frac{2}{(\gamma + 1){\cal M}^2}  \,,
\end{equation}
\begin{equation}\label{eq:rankine_2}
  p_{\textrm s}   =  \frac{2}{\gamma + 1} -
           \frac{\gamma - 1}{\gamma(\gamma + 1){\cal M}^2}  \,,
\end{equation}
where quantities immediately behind the shock are denoted with the 
subscript ${\textrm s}$, and $\gamma = 5/3$ will be used in what follows.

%%%%%%%%%%%%%%%%%%%%%%%%%%%%%%%%%%%%%%%%%%%%%%%%%%%%%%%%%%%%%%%%%%%%%%
\section{LINEAR THEORY}\label{sec:linear}
%
%
%
%
%
%%%%%%%%%%%%%%%%%%%%%%%%%%%%%%%%%%%%%%%%%%%%%%%%%%%%%%%%%%%%%%%%%%%%%%

Equilibrium configurations of radiative shock waves may be thermally
unstable. The nature of the instability may be interpreted as follows.

Consider a stationary shock, initially in equilibrium; 
if, say, the postshock temperature is slightly increased,
a longer cooling path will results and the excess pressure 
will force the shock to move upward. In the frame of the 
shock, the velocity of the incoming gas will increase even further
and the postshock temperature will rise according to the square 
of the preshock velocity.
If radiative losses are described by a decreasing function of the 
temperature, the cooling time will increase and the shock 
will continue to move upward.

%%%%%%%%%%%%%%%%%%%%%%%%%%%%%%%%%%%%%%%%%%%%%%%%%%%%%%%%%%%%%%%%%%%%%%
\subsection{Perturbative Analysis}\label{sec:perturb}
%
%
%
%%%%%%%%%%%%%%%%%%%%%%%%%%%%%%%%%%%%%%%%%%%%%%%%%%%%%%%%%%%%%%%%%%%%%%

A perturbative study is carried out by properly linearizing  
equations (\ref{eq:continuity})--(\ref{eq:energy})
around the steady-state solutions denoted with 
$\rho_0$, $v_0$ and $p_0$. The perturbed location of the shock 
front is written as
\begin{equation}
 x_{\textrm s} = 1 + \frac{\epsilon}{\delta}e^{\delta t} \,,
\end{equation}
where $x_{\textrm s} = 1$ is the shock equilibrium position in absence of
perturbation ($\epsilon = 0$), $\epsilon$ is the magnitude
of the perturbation, and $\delta$ is a complex eigenfrequency. 
According to the normalization scales introduced in \S\ref{sec:eqns},
time is expressed in units of $L_{\textrm c}/|v_{\textrm{in}}|$.

Following the same notations as in \citet{Sax98}, it is convenient
to write $\delta = \delta_{\textrm R} + i\delta_{\textrm I}$, where 
the real part $\delta_{\textrm R}$ gives the growth/decay term, 
while $\delta_{\textrm I}$ represents the oscillation frequency.
The nature of the instability is determined by the sign of $\delta_{\textrm R}$:
modes with negative $\delta_{\textrm R}$ are stable, while modes with 
positive $\delta_{\textrm R}$ are unstable.

Perturbed physical variables take the form 
\begin{equation}
  q(\xi,t) = q_0(\xi)\left(1 + \lambda_q(\xi) \epsilon e^{\delta t}\right) \,,
\end{equation}
where $q\in\{\rho,v,p\}$, $q_0(\xi)$ is the corresponding steady-state
value and the complex function $\lambda_q(\xi)$ describes the effects
of the perturbation. Here $\xi$ is a spatial coordinate normalized so that 
$\xi = 1$ at the shock and $\xi = 0$ at the wall:
\begin{equation}
 \xi = \frac{x}{x_{\textrm s}} \approx  x\left(1 - \frac{\epsilon}{\delta}e^{\delta t}\right) 
             + O(\epsilon^2) \,.
\end{equation}
 
The fluid equations are linearized in a frame of reference which is 
co-moving with the shock; in this frame the derivatives of a flow 
variable become
\begin{equation}
 \pd{}{t} \; \rightarrow \; \pd{}{t} + \pd{\xi}{t}\pd{}{\xi} \,, \quad
 \pd{}{x} \; \rightarrow \; \pd{\xi}{x}\pd{}{\xi}  \,.
\end{equation}

Therefore, retaining only terms up to first order in $\epsilon$,
one has 
\begin{equation}
  \pd{q}{t} \approx \Big( q_0 \lambda_q \delta - \xi q_0' \Big)
                \epsilon e^{\delta t} \,,
\end{equation}
\begin{equation}
  \pd{q}{x} \approx q_0' + \left( q'_0 \lambda_q + q_0\lambda_q' 
               - \frac{q_0'}{\delta}\right)\epsilon e^{\delta t} \,,
\end{equation}
where a primed quantity denotes a derivative with respect
to $\xi$.

The steady-state equations are obtained by collecting the
zeroth order terms in the Euler equations;
conservation of mass and momentum is trivially expressed by
\begin{equation}\label{eq:steady_1}
  \rho_0 v_0 = -1    \,,
\end{equation}
\begin{equation}\label{eq:steady_2}
   -v_0 + p_0 = m    \,,
\end{equation}
where the integration constants on the right hand sides may be evaluated
from the preshock values; hence $m = 1 + 1/(\gamma{\cal M}^2)$.

The pressure equation provides the explicit dependence
on the spatial coordinate; it can be put in closed 
integral form by writing
\begin{equation}\label{eq:steady_3}
 \xi(v_0) = \frac{f(v_0)}{f(v_{\textrm s})}\,,
\end{equation}
where 
\begin{equation}\label{eq:steady_int}
  f(v) = \int_0^v \left(-y\right)^{2-\alpha}
         \frac{\left[y  + \gamma(m + y)\right]}{(m + y)^\alpha}\,dy  \,,
\end{equation}
and $v_{\textrm s} = -(1 + 3/{\cal M}^2)/4$ is the fluid velocity 
immediately behind the shock (eq. [\ref{eq:rankine_1}]).
Notice that, according to the normalization units introduced in
\S\ref{sec:eqns}, the constant ${\cal C}$ in equation 
(\ref{eq:normalization}) takes the value
\begin{equation}\label{eq:constant_C}
 {\cal C} = -\frac{f(v_{\textrm s})}{(\gamma-1)}\,.
\end{equation}

Results pertinent to this section are evaluated in the strong shock
limit, ${\cal M}\, \rightarrow\, \infty$, so $m = 1$, 
$v_{\textrm s} = -1/4$ and $\alpha$ becomes the only free parameter in the problem.

\begin{figure}
%\epsscale{.80}
\plotone{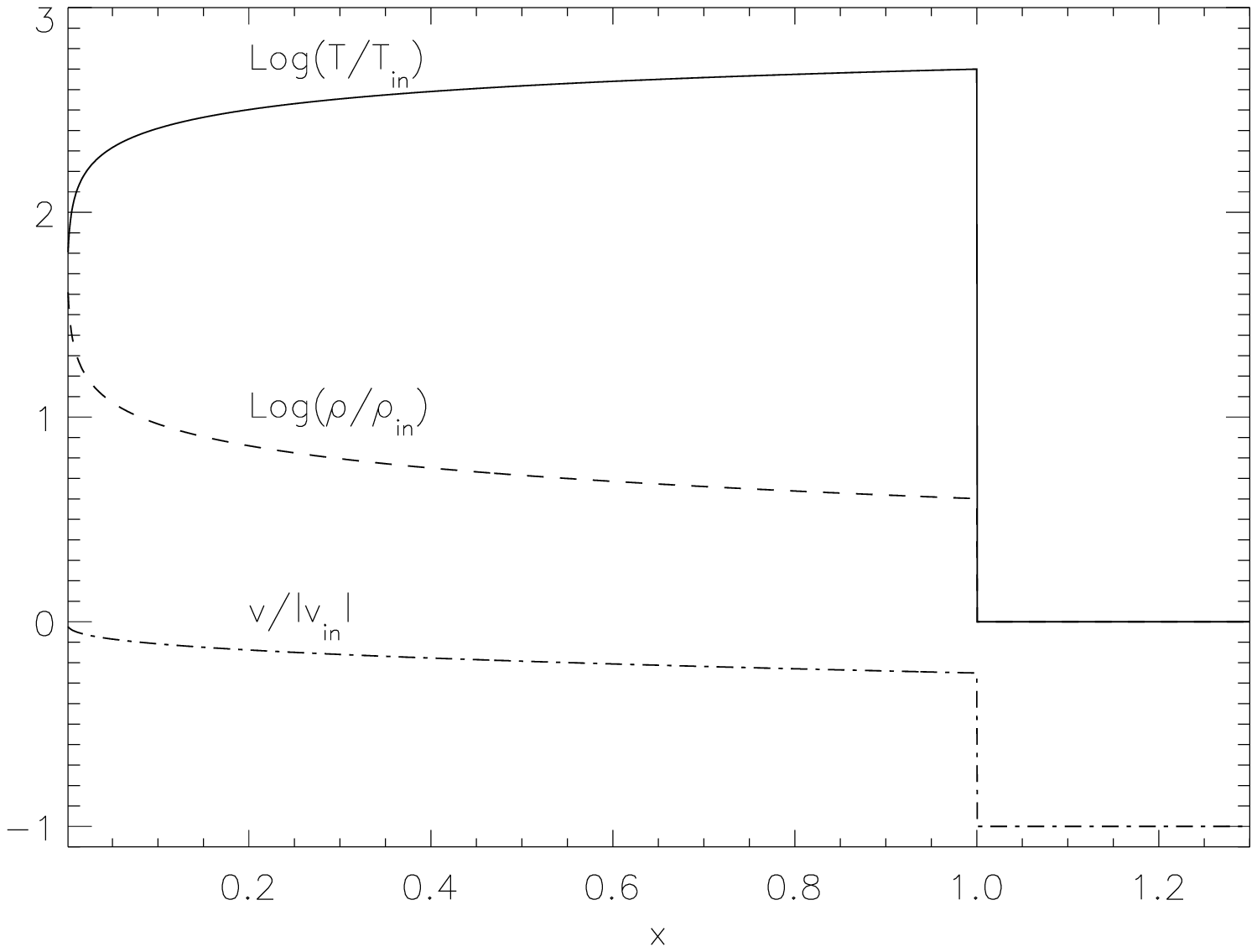}
\caption{\footnotesize Steady-state profiles for density, 
         temperature and velocity when $\alpha = 0$.
         The ``wall'' is located at $x=0$ and supersonic gas
         flows from the right to the left.
         Flow variables are normalized to their inflow values,
         and the abscissa is expressed in units of shock height.
         \label{fig:steady}}
\end{figure}

The integral in equation (\ref{eq:steady_int}) can be evaluated 
analytically for some specific values of the cooling index $\alpha$
(CI) but it has to be computed by numerical quadrature
in the general case. Notice that a steady-state solution is possible
only if the integral converges, that is, if $\alpha < 3$.
Equation (\ref{eq:steady_3}) can be inverted numerically to express 
the postshock steady flow velocity $v_0$ as a function of $\xi$.
The steady-state profiles are shown in Figure \ref{fig:steady}.

The first-order terms in $\epsilon$ provide three coupled complex 
differential equations for the perturbations; using
the unperturbed postshock velocity $v_0$ as the independent 
variable, they are
\begin{equation}\label{eq:pert_1}
   \frac{d\lambda_\rho}{dv_0} + \frac{d\lambda_v}{dv_0} =
    -\frac{\xi}{v_0^2}
    - \frac{\lambda_\rho\delta}{v_0}\frac{d\xi}{dv_0}  \,,
\end{equation}
\begin{equation}\label{eq:pert_2}
 v_0 \frac{d\lambda_v}{dv_0} - p_0\frac{d\lambda_p}{dv_0} = 
  - \lambda_v\delta\frac{d\xi}{dv_0}
  + \frac{\xi}{v_0} + \lambda_p - 2\lambda_v - \lambda_\rho  \,,
\end{equation}
\begin{equation}\label{eq:pert_3}
  v_0p_0\left(\gamma \frac{d\lambda_v}{dv_0} + 
                     \frac{d\lambda_p}{dv_0}\right) = 
       \left(v_0 + \gamma p_0\right) 
  \left[ (2-\alpha)\lambda_\rho + (\alpha - 1)\lambda_p
         - \lambda_v + \frac{1}{\delta}\right] 
       - p_0\lambda_p\delta \frac{d\xi}{dv_0}  + \xi \,,
\end{equation}
where $d\xi/dv_0$ is given by straightforward differentiation
of equation (\ref{eq:steady_3}) together with equation 
(\ref{eq:steady_int}).

For a given value of $\alpha$, equations (\ref{eq:pert_1}) 
through (\ref{eq:pert_3}) have to be solved by integrating
from the shock front, where $v_0 = v_{\textrm s}$, to the wall, 
where $v_0 = 0$.
The eigenmodes of the system are determined by imposing appropriate
boundary conditions to select the physically relevant solutions.
At the shock front the jump conditions for a strong shock 
(${\cal M}\to\infty$) apply \citep{Ima96,Sax98}:
\begin{equation}
   \lambda_\rho =  0 \,, \quad
   \lambda_v    = -3 \,, \quad
   \lambda_p    =  2 \,.
\end{equation}

At the bottom of the postshock region ($\xi=0$) the relevant
physical solutions must satisfy the stationary wall condition, 
namely, that the flow comes to rest and the velocity must be 
oscillation-free. This requires that both the real and imaginary
parts of $\lambda_v(v_0)$ vanish at $v_0=0$.
The complex frequencies $\delta$ for which such solutions
are possible identify the eigenmodes of the system.

The method of solution adopted here consists in minimizing 
the real function of two variables $\eta(\delta_{\textrm R}, \delta_{\textrm I}) = 
|\lambda_v(0)|$. Here $|\lambda_v(0)| = \left(\lambda_{v,\textrm{R}}^2(0) + 
\lambda_{v,\textrm{I}}^2(0)\right)^{1/2}$ is the magnitude of the velocity 
perturbation at the bottom of the postshock region; the values of 
the real and imaginary parts, $\lambda_{v,\textrm{R}}(0)$ and 
$\lambda_{v,\textrm{I}}(0)$, are obtained by direct numerical integration
of equations (\ref{eq:pert_1})--(\ref{eq:pert_3}) for a given pair  
$(\delta_{\textrm R},\delta_{\textrm I})$.
In practice, since the system is singular at the origin, integration 
proceeds from the shock up to some small value of $v_0$, denoted
with $v_\epsilon$. Setting $v_\epsilon \lesssim  10^{-3}$
did not produce significant variations in the solution.

\begin{figure}
%\epsscale{.80}
\plotone{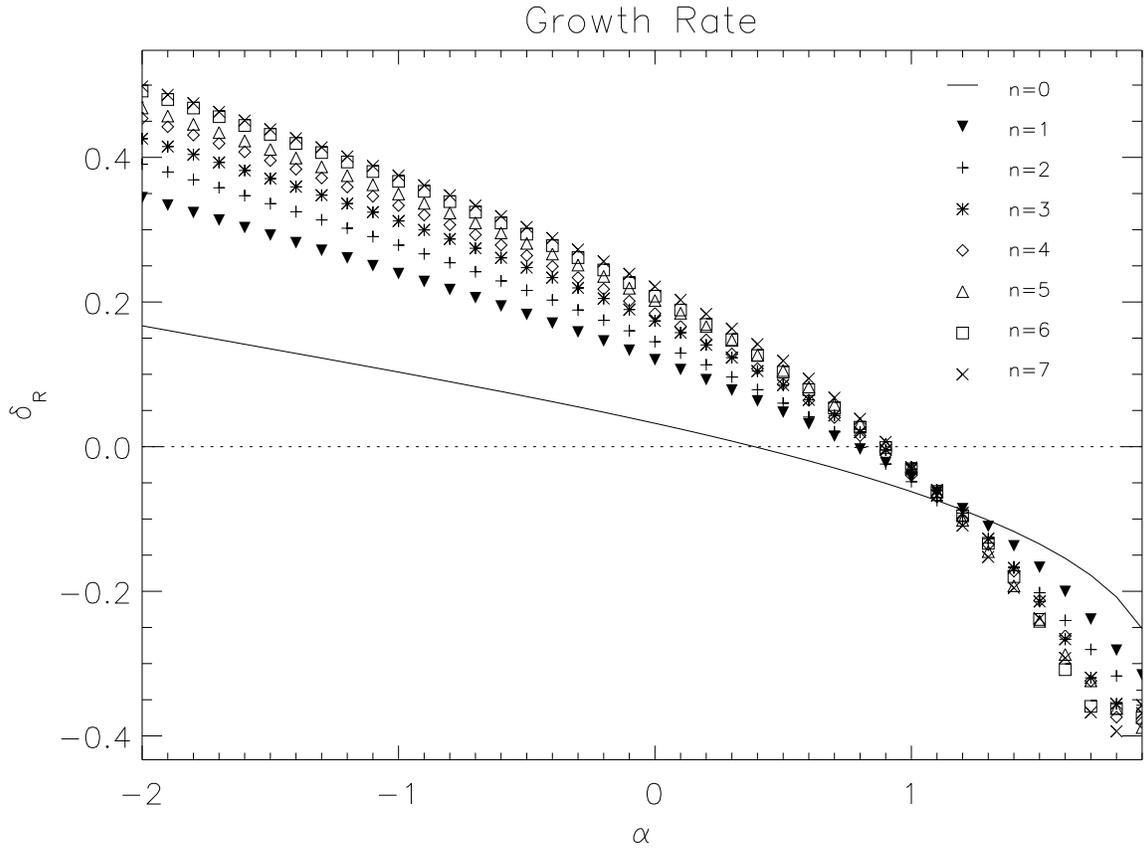}
\caption{\footnotesize Growth/damping rates for the
         first 8 modes as function of $\alpha$. 
         The solid line represents the fundamental mode $n = 0$,
         whereas the different symbols (described by the legend in
         the upper-right portion of the plot) correspond to the seven
         overtones  $1 \le n \le 7$. 
         Eigenmodes with $\delta_{\textrm R} < 0$ are stable, whereas modes with 
         $\delta_{\textrm R}>0$ are unstable. \label{fig:modes_re}}
\end{figure}

\begin{figure}
%\epsscale{.80}
\plotone{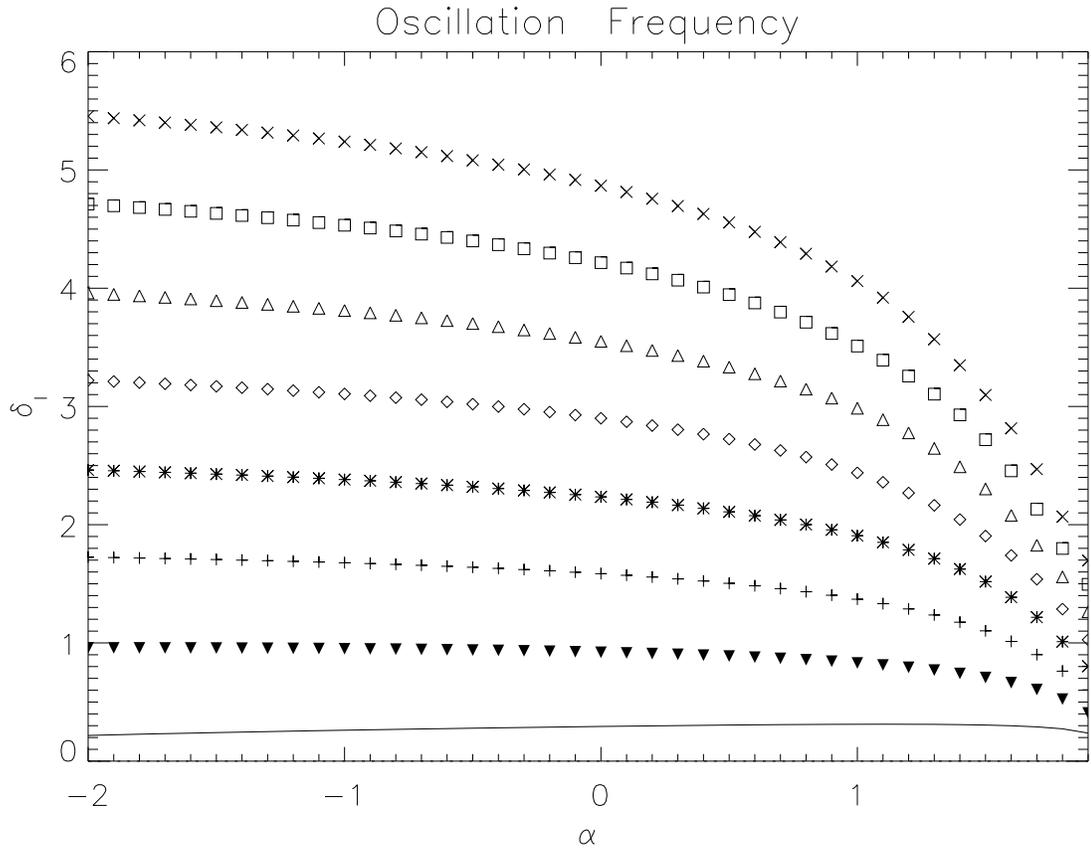}
\caption{\footnotesize Oscillation frequencies for the
         first 8 modes as function of $\alpha$. The symbols
         have the same meaning as in Figure \ref{fig:modes_re}.
         Modes with $1\le n\le7$ have oscillation frequencies that are
         monotonically decreasing functions of $\alpha$. For the 
         fundamental mode ($n=0$), however, $\delta_{\textrm I}$ increases to 
         reach a maximum at approximately $\alpha = 1.1$ and 
         decreases afterwards. \label{fig:modes_im}}
\end{figure}

A preliminary coarse search with trial values of $\delta_{\textrm R}$ 
and $\delta_{\textrm I}$ shows that, for a given value of $\alpha$,
an indefinitely long series of modes exists. 
Following CI, modes are labeled by increasing oscillation frequency,
so that $n=0$ correspond to the fundamental mode, $n=1$ to the
first overtone, $n=2$ to the second overtone, and so on. 
The approximate position of each mode $n$, $(\delta_{\textrm R}^{(n)},
\delta_{\textrm I}^{(n)})$, is then iteratively improved by repeating the search
on finer sub-grids (in the complex $\delta$ plane) centered around
the most recent iteration of $\delta^{(n)}_{\textrm R}$, $\delta^{(n)}_{\textrm I}$.
The process stops once the relative error between two 
consecutive iterations falls below $10^{-6}$.
For practical reasons, the search algorithm has been restricted to 
the first eight modes for values of $\alpha$ uniformly distributed 
in the range $-2 \le \alpha < 2$.
Results are shown in Figures \ref{fig:modes_re} and \ref{fig:modes_im}, 
while modes for some specific values of $\alpha$ are listed in Tables
\ref{tab:modes_1} and \ref{tab:modes_2}.

A mode is stable if the real part of the corresponding eigenvalue has
negative sign, and unstable otherwise.
High-frequency modes are characterized by growth rates 
which decrease faster than low-frequency ones for increasing $\alpha$.
Hence, the fundamental mode ($n=0$) has the smallest growth/damping
rate for $\alpha \lesssim 1$, but the smallest damping rate
for $\alpha \gtrsim 1$.
Modes with $n\ge 1$ have monotonically decreasing oscillation 
frequencies while, for the fundamental mode, $\delta_{\textrm I}$ reaches a 
maximum value at $\alpha\approx 1.1$ and decreases afterwards.

%%%%%%%%%%%%%%%%%%%%%%%%%%%%%%%%%%%%%%%%%%%%%%%%%%%%%%%%%%%%%%%%%%%%%%
\subsection{Critical $\alpha$}\label{sec:critical}
%
%
%
%%%%%%%%%%%%%%%%%%%%%%%%%%%%%%%%%%%%%%%%%%%%%%%%%%%%%%%%%%%%%%%%%%%%%%
 
\begin{figure}
\plotone{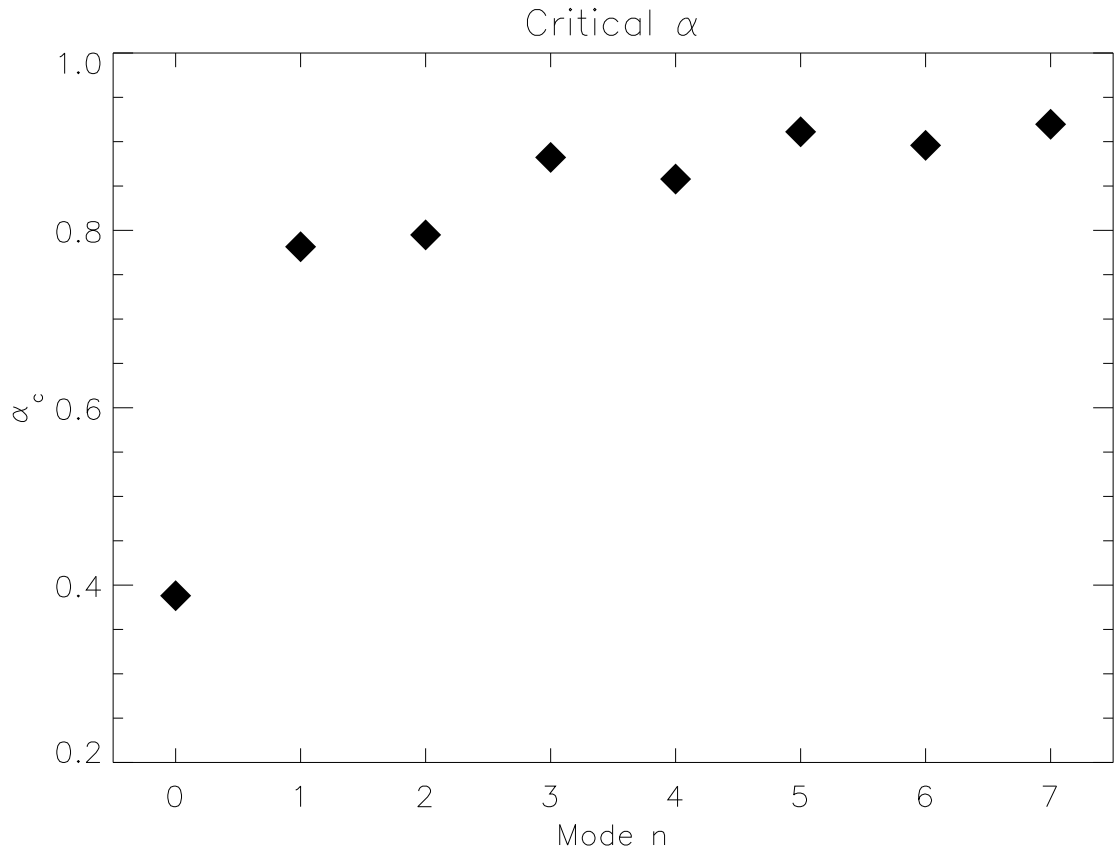}
\caption{\footnotesize Critical value of the cooling index
         as function of the mode number $n$.
         For a given mode $n$, values of $\alpha > \alpha_{\textrm c}$ have 
         negative growth rates and thus are stable. \label{fig:alpha_crit}}
\end{figure}

For each mode $n$, a critical value of the cooling index,
$\alpha_{\textrm c}^{(n)}$, may be defined, such that $\delta_{\textrm R}^{(n)} = 0$
when $\alpha = \alpha^{(n)}_{\textrm c}$.
Hence, the $n$-th mode is stable for $\alpha > \alpha^{(n)}_{\textrm c}$
and unstable when $\alpha < \alpha^{(n)}_{\textrm c}$ (Fig. \ref{fig:alpha_crit}).
The value of the critical $\alpha$ is computed by interpolating $\alpha$
with a quartic polynomial passing through the two pairs of values
across which $\delta_{\textrm R}$ changes sign.
Thus the fundamental mode becomes stable for $\alpha > 0.388$, the first
harmonic for $\alpha > 0.782$, and so on. Values of $\alpha_{\textrm c}$ are
listed in Table \ref{tab:modes_2} and shown in Figure \ref{fig:alpha_crit}.
Interestingly, the sequence of critical $\alpha$ is not monotonic 
with increasing $n$.
Finally, notice that all (8) modes become eventually stable for 
$\alpha \gtrsim 0.92$.

%%%%%%%%%%%%%%%%%%%%%%%%%%%%%%%%%%%%%%%%%%%%%%%%%%%%%%%%%%%%%%%%%%%%%%
\subsection{Linear Fit}\label{sec:fit}
%
%
%
%%%%%%%%%%%%%%%%%%%%%%%%%%%%%%%%%%%%%%%%%%%%%%%%%%%%%%%%%%%%%%%%%%%%%%

By inspecting Figure \ref{fig:modes_im}, one can easily see
that, for a given $\alpha$, the oscillation frequencies 
of the different modes are approximately equally spaced as
$n$ increases.
In this respect, they resembles the modal frequencies of  
a pipe open at one end \citep{TD93, Sax98, Sax99} and can
be described by a simple linear fit of the form
\begin{equation}\label{eq:linfit}
 \delta^{(n)}_{\textrm I} = \tilde{\delta}^{(0)}_{\textrm I} 
                           + n \Delta\tilde{\delta}_{\textrm I} \,.
\end{equation}
with a small residual, $\lesssim 0.5\%$.
In equation (\ref{eq:linfit}), $\tilde{\delta}^{(0)}_{\textrm I}$ is the
``fitted'' fundamental frequency and $\Delta\tilde{\delta}_{\textrm I}$
is a frequency spacing depending on the cooling index $\alpha$.
$\Delta\tilde{\delta}_{\textrm I}$ is monotonically decreasing
for increasing $\alpha$. Values of $\tilde{\delta}_{\textrm I}^{(0)}$ and
$\Delta\tilde{\delta}_{\textrm I}$ are given in Tables \ref{tab:modes_1}
and \ref{tab:modes_2}.

%%%%%%%%%%%%%%%%%%%%%%%%%%%%%%%%%%%%%%%%%%%%%%%%%%%%%%%%%%%%%%%%%%%%%%
\section{TIME-DEPENDENT NUMERICAL SIMULATIONS}\label{sec:time_dependent}
%
%
%
%
%
%%%%%%%%%%%%%%%%%%%%%%%%%%%%%%%%%%%%%%%%%%%%%%%%%%%%%%%%%%%%%%%%%%%%%%

The results of the previous sections indicate that radiative
shocks in real astrophysical settings may be linearly unstable
and thus far from an equilibrium configuration.
This calls for the investigation of the full time-dependent
problem where nonlinear effects may play a major role in the shock
dynamics.

In what follows, the radiative shock evolution is analyzed through a set 
of numerical simulations for different values of the cooling index $\alpha$.
The early evolutionary phases are of particular interest since they can 
be directly compared to the expected linear behavior, thereby providing
an effective tool in validating the correctness of the numerical method
and choice of boundary conditions.
Nonlinear effects, on the other hand, describe the long-term dynamics
of the shock and play a crucial role in determining whether
a linearly stable mode may actually be nonlinearly unstable \citep{Sax99}.

%%%%%%%%%%%%%%%%%%%%%%%%%%%%%%%%%%%%%%%%%%%%%%%%%%%%%%%%%%%%%%%%%%%%%%
\subsection{Numerical Method}\label{sec:num_meth}
%
%
%
%%%%%%%%%%%%%%%%%%%%%%%%%%%%%%%%%%%%%%%%%%%%%%%%%%%%%%%%%%%%%%%%%%%%%%
 
The numerical approach followed here relies on the high-resolution
shock-capturing methods (HRSC henceforth, see \cite{LeVeque98}
and references therein).
HRSC methods rely on a finite-volume, conservative discretization
of the Euler equations, thus being particularly suitable in 
describing shock dynamics and, in general, modeling
flow discontinuities.

The starting point is the system of equations 
(\ref{eq:continuity}) through (\ref{eq:energy})
written in conservative form:
\begin{equation}\label{eq:cons_equations}
  \pd{\A{U}}{t} = -\pd{\A{F}}{x} + \A{S} \,,
\end{equation}
where $\A{U} = \big(\rho, \rho v, E\big)$ is a vector
of conservative quantities, while 
\begin{equation}
 \A{F} = \left(\begin{array}{c}
  \DS \rho v  \\        \noalign{\medskip}
  \DS \rho v^2 + p \\    \noalign{\medskip}
  \DS (E + p)v    \\
\end{array}\right)  \,, \quad
 \A{S} = \left(\begin{array}{c}
  \DS  0  \\        \noalign{\medskip}
  \DS  0  \\    \noalign{\medskip}
  \DS -{\cal C}\rho^{2-\alpha}p^\alpha   \\    
\end{array}\right)  \,,
\end{equation}
are the flux and source term vectors, respectively.
The  total energy density $E$ is expressed as the sum
of internal and kinetic terms:
\begin{equation}\label{eq:eos}
  E = \frac{p}{\gamma - 1} + \frac{\rho v^2}{2} \,.
\end{equation}

The system of equations (\ref{eq:cons_equations})
is solved numerically using {\it PLUTO}, a high-resolution
Godunov-type code for astrophysical fluid dynamics
(Mignone et al. 2005, in preparation).

With {\it PLUTO}, equations (\ref{eq:cons_equations}) are solved
by operator splitting, i.e., by treating the advection term
$\partial\A{F}/\partial x$ and the source term $\A{S}$ in
separate steps.
This approach is second order accurate in time
if the two operators have at least the same accuracy 
and the order into which they are applied is reversed 
every time step \citep{Strang68}.

During the advection step, a high-resolution, shock capturing 
Godunov-type  formulation is adopted.
Second-order accuracy in space is based on a conservative, 
monotonic interpolation of the characteristic fields
\citep{Colella90}.
Third-order temporal accuracy is achieved by a 
multi-step Runge-Kutta TVD algorithm \citep{Gottlieb_Shu98}:
\begin{equation}
 \begin{array}{c}
 \DS \A{U}^1_j = \A{U}_j^n + \Delta t^n \tilde{\A{R}}_j(\A{U}^n)    \,, \\ \noalign{\medskip}
 \DS \A{U}^2_j = \frac{1}{4}\left(3\A{U}_j^n + \A{U}_j^1 
                     + \Delta t^n \tilde{\A{R}}_j(\A{U}^1)\right)   \,, \\ \noalign{\medskip}
 \DS \A{U}^{n+1}_j = \frac{1}{3}\left(\A{U}_j^n + 2\A{U}_j^2 
                     + 2\Delta t^n \tilde{\A{R}}_j(\A{U}^2)\right)  \,. 
 \end{array} 
\end{equation}
Here $\tilde{\A{R}}_j(\A{U})$ is a conservative, discretized approximation 
to the flux term on the right hand side of equation 
(\ref{eq:cons_equations}):
\begin{equation}
 \tilde{\A{R}}_j(\A{U}) = -\frac{\tilde{\A{F}}_{j+\HALF} 
                          - \tilde{\A{F}}_{j-\HALF}}{\Delta x_j}  \,,
\end{equation}
where $j$ labels the computational zone with mesh spacing $\Delta x_j$. 
The numerical fluxes $\tilde{\A{F}}_{j\pm\HALF}$ are computed using 
the approximate Riemann solver of \citet{Roe81}.
It should be mentioned that, although different numerical schemes have
also been employed, no significant differences were found in the results
presented in \S\ref{sec:results}. 
The particular choice of Riemann solver and interpolation algorithm
is quite common and represents a good trade-off between accuracy and
computational time.

Cooling is treated in a separate source step, where 
\begin{equation}\label{eq:cooling_step}
   \frac{dE}{dt} = -{\cal C} \rho^{2-\alpha} p^\alpha  \,,
\end{equation}
with ${\cal C}$ given by equation (\ref{eq:constant_C}),
is solved.
Notice that only the internal energy is affected during
the source step, while density and velocity remain
constant with the values provided by the most recent advection
step.
Thus, the kinetic contribution to the total energy can be discarded 
and equation (\ref{eq:cooling_step}) provides an ordinary 
differential equation in the pressure variable only.
Integration for a time step $\Delta t^n$ can be done 
analytically:
\begin{equation}\label{eq:pressure_int}
   p^{n+1} = \left\{\begin{array}{cc}
  \DS \left[p^{1 - \alpha} - \Delta t^n {\cal C}(\gamma-1)
            \rho^{2-\alpha}
           (1-\alpha)\right]^{\frac{1}{1-\alpha}} 
     &  \quad \textrm{if} \quad \alpha \neq 1  \,, \\ \noalign{\medskip}
  \DS p \exp\left(-{\cal C}(\gamma -1)\Delta t^n 
            \rho^{2-\alpha}\right) 
     & \quad  \textrm{if} \quad \alpha = 1  \,,
   \end{array}\right.  
\end{equation}
where $p$ and $\rho$ are the pressure and density at the beginning 
of the source step and the suffix $j$ has been omitted for 
clarity of exposition.

Radiative losses are identically zero for $T < T_{\textrm c}$, where
$T = p/\rho$ is a dimensionless temperature and $T_{\textrm c}$ is the cutoff
temperature, equal to the temperature of the incoming gas, i.e., 
$T_{\textrm c} = 1/(\gamma{\cal M}^2)$.

%%%%%%%%%%%%%%%%%%%%%%%%%%%%%%%%%%%%%%%%%%%%%%%%%%%%%%%%%%%%%%%%%%%%%%
\subsection{Initial and Boundary conditions}\label{sec:boundary}
%
%
%
%%%%%%%%%%%%%%%%%%%%%%%%%%%%%%%%%%%%%%%%%%%%%%%%%%%%%%%%%%%%%%%%%%%%%%

The computational domain is the region $x_0 \le x \le x_1$, where
$x_0$ and $x_1$ define the locations of the lower and
upper boundaries, respectively. As in the previous sections, 
lengths are expressed in units of the cooling thickness in steady-state;
hence the shock is initially located at $x = 1$.

Upstream, for $1 < x < x_1$, density, velocity 
and pressure are uniform and equal to the preshock values, i.e. 
$\rho = 1$, $v = -1$ and $p= 1/(\gamma{\cal M}^2)$. At $x = x_1$ a 
constant supersonic inflow defines the upstream boundary condition. 
Here $x_1 = 5$ and ${\cal M} = 40$ will be used for all simulations.
Downstream, for $x_0 \le x \le 1$, flow quantities are initialized 
to the steady-state solution given by equations (\ref{eq:steady_1})
through (\ref{eq:steady_3}). 
The location of the lower boundary $x_0$ depends on the particular
boundary condition adopted.
 
The implementation of a suitable boundary condition at the lower 
boundary poses a serious nontrivial problem for a number of different 
reasons.  
Close to the wall, density and velocity experience rapidly varying 
sharp gradients, thus demanding increasingly high resolution in order to
resolve flow patterns. This, in turns, has the inevitable consequence
of restricting the time step size in an explicit code.
Although a number of different strategies have been proposed,
there is no general agreement about what would be a ``consistent''
boundary condition. A comparison between four different approaches
is considered in this paper and results are discussed in detail
in the next sections.

A first approach consists in using a reflecting boundary 
condition (RFbc), commonly adopted \citep[][SB]{Plewa95} to simulate 
the presence of a rigid wall or to enforce axial or planar symmetry. 
It imposes symmetric profiles (with respect to the ``wall'' position $x_0 = 0$)
on density and pressure, while the velocity is antisymmetric, i.e.,
$v(- x) = -v(x)$.  
Hence at the lower boundary, the velocity is zero at all times while density
and pressure have zero gradient. 

A second approach adopts a ``fixed'', time-independent 
boundary conditions (FXbc) where flow variables are kept constant
at the steady-state values at $x = x_0$. In this case, $x_0 = 10^{-3}$ 
can be safely used, since is put sufficiently close to
the wall but such that the temperature at that point is still 
above the cutoff temperature. The cutoff temperature, 
therefore, has only the effect of preventing cooling in the
upstream region.  

I also introduce a new, third approach based on the characteristic 
boundary method \citep{Thompson87,Thompson90}, the details of which 
are given in Appendix \ref{app:cvbc}.
Following the same approach used in \S\ref{sec:perturb},
the velocity will be held constant at the steady-state value, whereas
pressure and density are allowed to evolve with time.
The ``constant-velocity'' boundary condition is hereafter referred to as 
CVbc.
 
Finally, a fourth recipe (SB, SDB, LCSb) is to further extend the domain 
in the downstream direction by placing a cold dense layer 
for $x_{\textrm{low}} < x < x_0$. Here $x_{\textrm{low}} = -2$ is the new location of the
lower boundary, while $x_0 = 10^{-4}$ still lies inside the postshock region.
Flow quantities have constant profiles, continuous at $x = x_0$ and
the temperature of the cold layer becomes approximately 
$2\%$ of the temperature immediately behind the shock.
For this boundary condition, the cutoff temperature was set equal 
to the temperature of the cold dense layer.
At the back of the layer, $x = x_{\textrm{low}}$, an outflow boundary 
condition (OBbc) holds on density, velocity and pressure. 

The onset of instability is triggered by the discretization
error of the numerical scheme and no external, ``ad hoc''
perturbation is introduced, unless otherwise stated.

A uniform grid is used in the region $x_0\le x \le 1.4$, 
whereas a second, geometrically stretched grid covers the rest of
the upstream region, $1.4 < x \le 5$.
The extent of the uniform region has been chosen to ensure that 
the largest-amplitude oscillations would adequately be resolved.

Issues concerning grid resolution effects must not be 
underestimated. In fact, as outlined by \citet[SBD hereafter]{SBD03},
sharp density gradients can be described with relatively 
limited accuracy because of numerical diffusion effects that
cause high-density regions to ``leak'' mass into neighboring 
low-density zones.
Since the cooling process is proportional to the square of density,
radiative losses will generally be overestimated, causing 
abnormal, excessive cooling.
Although this issue is intrinsic to any grid of finite size and
cannot be completely removed, higher resolutions can considerably
mitigate the problem.
Furthermore, small-amplitude oscillations of the shock
front can be adequately captured on finer grids.

For this reason, a grid of $2240$ computational zones 
covers the extent of the uniform region for all numerical
simulations presented in this paper. With this resolution,
the postshock flow is initially resolved on $1600$ points.
The number of points for the stretched grid has virtually no
influence on the solution and is held fixed at $200$ in
all cases.

%%%%%%%%%%%%%%%%%%%%%%%%%%%%%%%%%%%%%%%%%%%%%%%%%%%%%%%%%
\subsection{Results}\label{sec:results}
%
%
%
%
%%%%%%%%%%%%%%%%%%%%%%%%%%%%%%%%%%%%%%%%%%%%%%%%%%%%%%%%%

The one-dimensional numerical simulations are carried out 
for five different values of $\alpha$, selected according to 
the linear analysis results presented in \S\ref{sec:linear}.

I first consider the case where $\alpha = 0$, since all modes 
have positive (linear) growth rates.
Next, $\alpha = 0.5$ and $\alpha = 0.7$ are examined, since 
all the overtones save the fundamental have positive linear growth rates
and are, therefore, unstable.
The value $\alpha = 0.8$ lies just above the instability limit of  
the first and second harmonics; therefore, as one can see from Table
\ref{tab:modes_2}, only overtones with $n\ge 3$ are expected to be
unstable.
Finally, I consider the case $\alpha = 1$, for which all the first
eight modes have been shown to have negative growth rates and thus
are stable.

%%%%%%%%%%%%%%%%%%%%%%%%%%%%%%%%%%%%%%%%%%%%%%%%%%%%%%%%%
\subsubsection{$\alpha = 0.0$}
%
%
%%%%%%%%%%%%%%%%%%%%%%%%%%%%%%%%%%%%%%%%%%%%%%%%%%%%%%%%%

\begin{figure}
\plotone{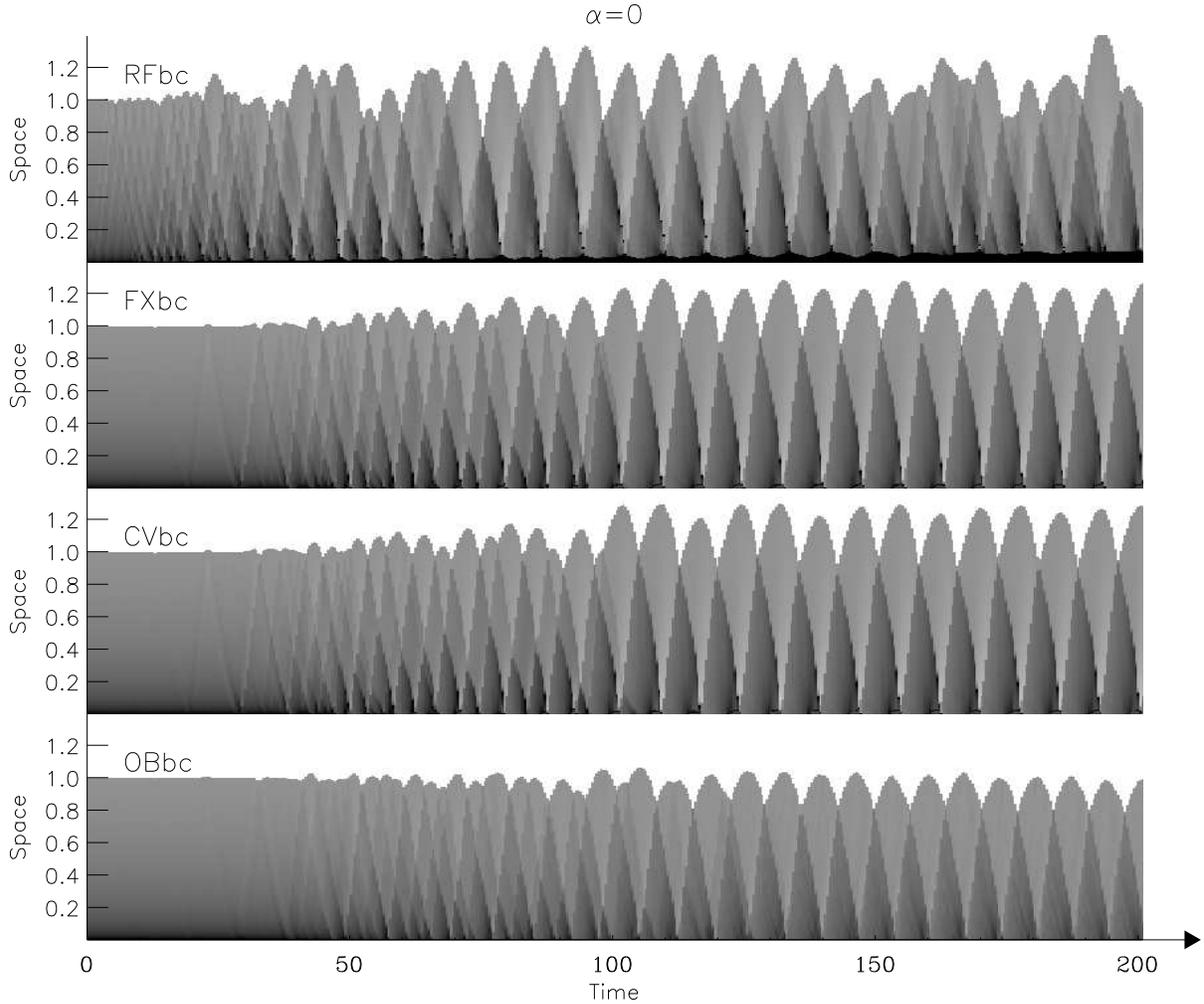}
\caption{\footnotesize Space-time diagrams for $\alpha = 0$ and the
         four different boundary conditions described in the text.
         From top to bottom: the reflective boundary 
         condition (RFbc), the fixed boundary condition (FXbc), 
         the constant velocity boundary condition (CVbc) and 
         the ``open boundary'' condition (OBbc).
         The spatial coordinate is represented on the vertical
         axis, whereas the time evolution of the system is described
         by the horizontal axis.
         Time is expressed in units of $L_{\textrm c}/|v_{\textrm{in}}|$, 
         where $L_{\textrm c}$ is the
         length of the cooling region in steady-state, see \S\ref{sec:perturb},
         and $v_{\textrm{in}}$ is the fluid velocity ahead of the shock.
         In each panel, the gas flows supersonically from the top
         to the bottom. The gray-scale shows the density logarithm: 
         lighter (darker) shades corresponds to lower (higher) density
         regions.\label{fig:alpha00}}
\end{figure}

\begin{figure}
%\epsscale{.80}
\plotone{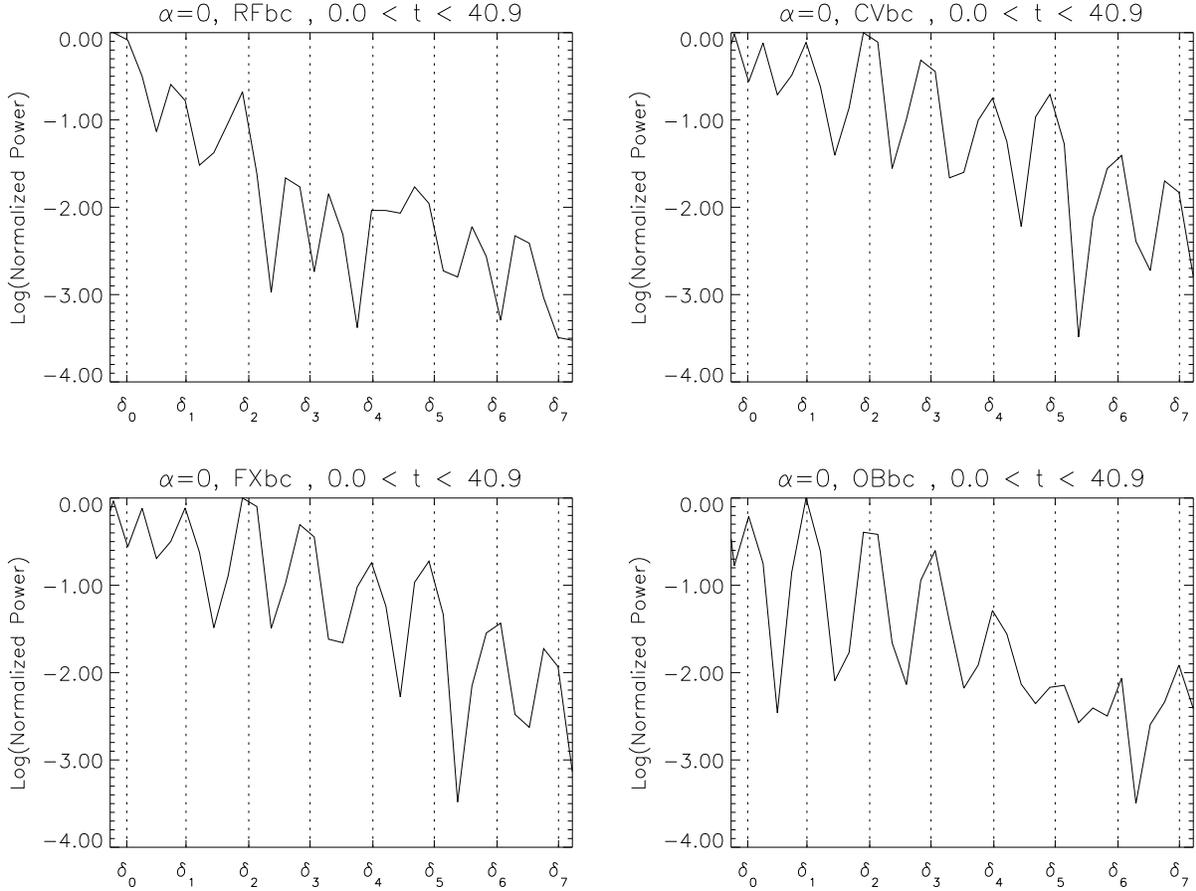}
\caption{\footnotesize Power spectra of the shock position for 
         $\alpha = 0.0$ derived from the numerical simulations 
         for the four boundary conditions described
         in the text. On the top: reflective (left) and 
         constant velocity boundary conditions (right); on the bottom:
         the fixed (left) and open boundary condition (right). 
         The vertical axis represents power on a logarithmic scale, 
         normalized to its maximum value. The horizontal axis shows 
         frequency on a linear scale. 
         The power spectra are obtained by computing the Fourier 
         transform of the function $x_{\textrm{sh}}(t) - x_{\textrm{sh}}(0)$, 
         where $x_{\textrm{sh}}(t)$ is the shock position at time $t$.
         The transform is taken over a sample of length $0 < t < 40.9$ 
         The dashed vertical lines in correspondence of the $\delta_n$,
         with $0\le n \le 7$, mark the oscillation eigenfrequencies
         derived from linear analysis, see Table \ref{tab:modes_1}.
         \label{fig:alpha00_pow}}
\end{figure}

The time-evolution diagrams for the four boundary 
conditions presented in \S\ref{sec:boundary} are shown 
in the four panels of Figure \ref{fig:alpha00}
and the corresponding power spectra of the shock oscillations
are shown in Figure \ref{fig:alpha00_pow}.

Significant departure from the steady-state solution occurs most rapidly 
when the RFbc is employed. In this case, the amplitude of the oscillations
rapidly increases with time and the system enters a nonlinear saturated
regime around $t \sim 20$ after a short-lived linear phase.
The use of the RFbc yields oscillation frequencies which are found to be 
shifted with respect to the values obtained from linear analysis.
This particular choice of boundary condition, in fact, forces the velocity
to have a node at the location of the lower boundary, while density and 
pressure have an extremum. 
Since this condition is far from the equilibrium configuration (eqns.
[\ref{eq:steady_1}]--[\ref{eq:steady_3}]), strong nonlinear perturbations
originate in the postshock flow and steepen into secondary shocks
at a high rate.
This also contributes to the higher amplitude oscillations observed
in this case.
Not surprisingly, the power spectrum of the shock position,
Figure \ref{fig:alpha00_pow}, exhibits frequencies that are
offset from the ones predicted by linear analysis.
Notice that, since the mass flux through the lower boundary
is zero and cooling is not effective for $T < T_{\textrm c}$, 
a cold layer of gas at $T = T_{\textrm c}$ accumulates at the 
bottom of the cooling region (Fig. \ref{fig:alpha00}).
Inside this layer, density is approximately constant and equal to 
$\gamma {\cal M}^2$, whereas waves propagate at the local sound 
speed, $c_s\sim 1/{\cal M}$.
 
In contrast, the CVbc and FXbc yield similar results and
the system preserves profiles close to the initial steady-state
values. The early phase of evolution ($t \lesssim 60\div 70$) 
is characterized by a linear growth of the perturbation, while,
for $t \gtrsim 80$, the amplitude of the oscillations begins to saturate
and the instability becomes nonlinear. During this phase
the largest oscillation amplitudes reach $\sim25\%$ of the initial
equilibrium position.
The OBbc shows reduced amplitudes with respect to the previous 
cases. In addition, the linear phase is longer than the CVbc or FXbc, and 
the transition to the nonlinear regime occurs only for $t \gtrsim 110$.

The power spectra for the early phase of the evolution
($0<t<41$) are plotted in Figure \ref{fig:alpha00_pow}.
Both the CVbc and FXbc yield eigenfrequencies that can be definitely
identified with the theoretical values, with a bigger uncertainty 
in the fundamental mode, see Table \ref{tab:error}.
Results obtained with the OBbc are similar and the fundamental
mode differs from the analytical expectation by less than $4\%$.

One should bear in mind, however, that the linear growth rate
of the fundamental mode is a factor of $\approx 6$ smaller than
those of higher harmonics ($1\le n\le 7$, see Table \ref{tab:modes_2}), 
and therefore modes with $n \ge 1$ tend to saturate faster.
For this reason, a representative sample of the linear phase must 
have a limited length in the time domain and, consequently, the 
power spectrum inevitably suffers from poor resolution at lower frequencies.

\begin{figure}
%\epsscale{.80}
\plotone{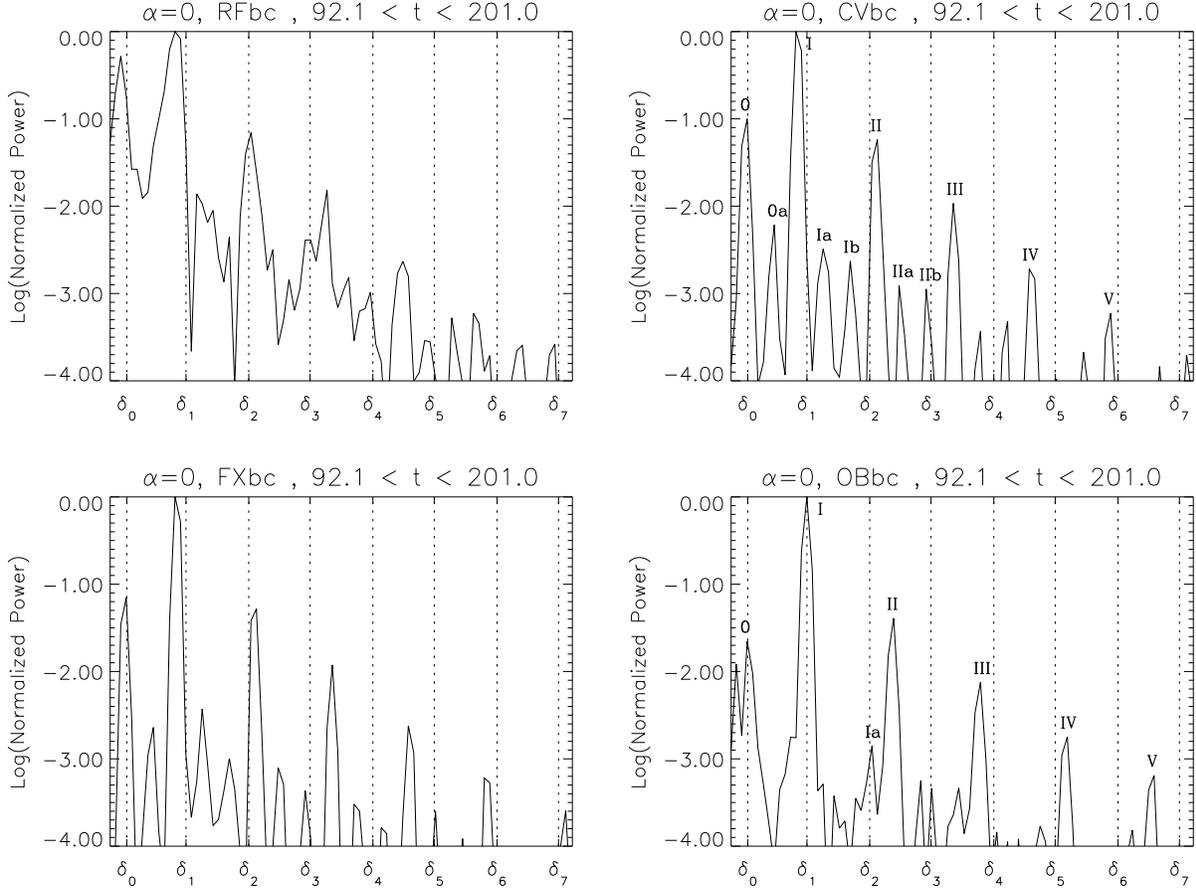}
\caption{\footnotesize Same as Figure \ref{fig:alpha00_pow}, but 
         when a longer portion of the shock position
         at later evolutionary times ($92.1 < t < 201$) is
         Fourier-transformed. For the CVbc and OBbc, a main sequence 
         of overtones ({\textrm I, II, III} and so forth) with frequencies 
         $\Omega^{\textrm{(I)}}$, $\Omega^{\textrm{(II)}}$, $\Omega^{\textrm{(III)}}$, \dots,
         may be identified from the plots. These modes 
         are almost equally spaced in frequency and may result  
         from nonlinear mode-mode coupling between the fundamental mode 
         and the first overtone.
         Secondary harmonics with oscillation frequencies
         $\Omega^{\textrm{(Ia)}}$, $\Omega^{\textrm{(IIa)}}$, $\Omega^{\textrm{(IIb)}}$, etc., 
         appear between the main sequence modes. 
         The explicit values of the $\Omega$'s are given 
         in Table \ref{tab:nonlinear}.\label{fig:alpha00_pow_late}}
\end{figure}

The situation is different when a longer portion 
($92 \lesssim t \lesssim 201$) of the shock position
during the saturated regime is analyzed, see Figure \ref{fig:alpha00_pow_late}.
In this case, the predominant mode of instability is the first overtone,
whereas the fundamental mode and the second harmonic contribute by
less than $10\%$ to the oscillatory cycle.
Notice also that when the CVbc and FXbc are used, the prevalent frequency of 
oscillation differs from the linear prediction by $\sim 10\%$, but it 
coincides with the first harmonic when the OBbc is adopted.

In all cases, a main sequence of harmonics with 
increasing frequencies $\Omega^{\textrm{(I)}}$, $\Omega^{\textrm{(II)}}$, 
$\Omega^{\textrm{(III)}}$, etc., 
may be identified by inspecting Figure \ref{fig:alpha00_pow_late}.
These overtones have monotonically decreasing power and  
do not necessarily coincide with the linear modes, but result 
from nonlinear interactions.
Nonlinearity enters through mode-mode coupling, as it
is suggested by considering the frequency spacing between them 
(Table \ref{tab:nonlinear}).
For the CVbc and OBbc, in fact, the spacing appears to be either a
multiple of the fundamental mode or equal to the first overtone. 
In the OBbc case, for instance, one has that 
$\Omega^{\textrm{(II)}} - \Omega^{\textrm{(I)}}  \approx 
 \Omega^{\textrm{(III)}} - \Omega^{\textrm{(II)}}
 \approx \Omega^{\textrm{(IV)}} - \Omega^{\textrm{(III)}} \approx 
 \Omega^{\textrm{(I)}}$, i.e., 
the mode spacing is a multiple of the first overtone.
A similar result holds in the CVbc case where, from Table 
\ref{tab:nonlinear}, it can be verified that
$\Omega^{\textrm{(II)}} - \Omega^{\textrm{(I)}}  \approx 3\Omega^{(0)}$, 
$\Omega^{\textrm{(III)}} - \Omega^{\textrm{(II)}} \approx 
 \Omega^{\textrm{(IV)}} - \Omega^{\textrm{(III)}} \approx \Omega^{\textrm{(I)}}$, and so 
on.

Intermediate, secondary peaks associated with small-power modes
appears between the main sequence overtones.
Some of these modes have been identified and labeled in 
Figure \ref{fig:alpha00_pow_late} with $\Omega^{\textrm{(Ia)}}$, 
$\Omega^{\textrm{(Ib)}}$, $\Omega^{\textrm{(IIa)}}$, $\Omega^{\textrm{(IIb)}}$, etc..
These secondary overtones may result from mode-mode coupling between 
the main sequence modes and the fundamental. This coupling is 
most evident for the CVbc, where one finds that $\Omega^{\textrm{(0a)}} - \Omega^{(0)} \approx 
\Omega^{\textrm{(Ia)}} - \Omega^{\textrm{(I)}} \approx \Omega^{\textrm{(II)}} - \Omega^{\textrm{(Ib)}}
\approx \Omega^{(0)}$, and similarly for higher harmonics 
(see Tab. \ref{tab:nonlinear}).
% ---------------------------------------------------------------------

%%%%%%%%%%%%%%%%%%%%%%%%%%%%%%%%%%%%%%%%%%%%%%%%%%%%%%%%%
\subsubsection{$\alpha = 0.5, 0.7$}\label{sec:alpha0507}
%
%
%%%%%%%%%%%%%%%%%%%%%%%%%%%%%%%%%%%%%%%%%%%%%%%%%%%%%%%%%

Based on the previous results and considerations, the CVbc, FXbc and 
OBbc yield results that more accurately reproduce the predicted 
linear behavior during the system's early phase of evolution. 
Moreover, results obtained with the FXbc and CVbc exhibits strong similarities,
and thus only the CVbc and OBbc will be considered in what follows.

The $\alpha = 0.5$ value is of particular astrophysical relevance, 
since it describes optically thin bremsstrahlung, which is the main source
of radiative losses at temperatures of the order of $10^8\div 10^9$ K, 
typical in accretion shocks in magnetic cataclysmic variables.

\begin{figure}
\plotone{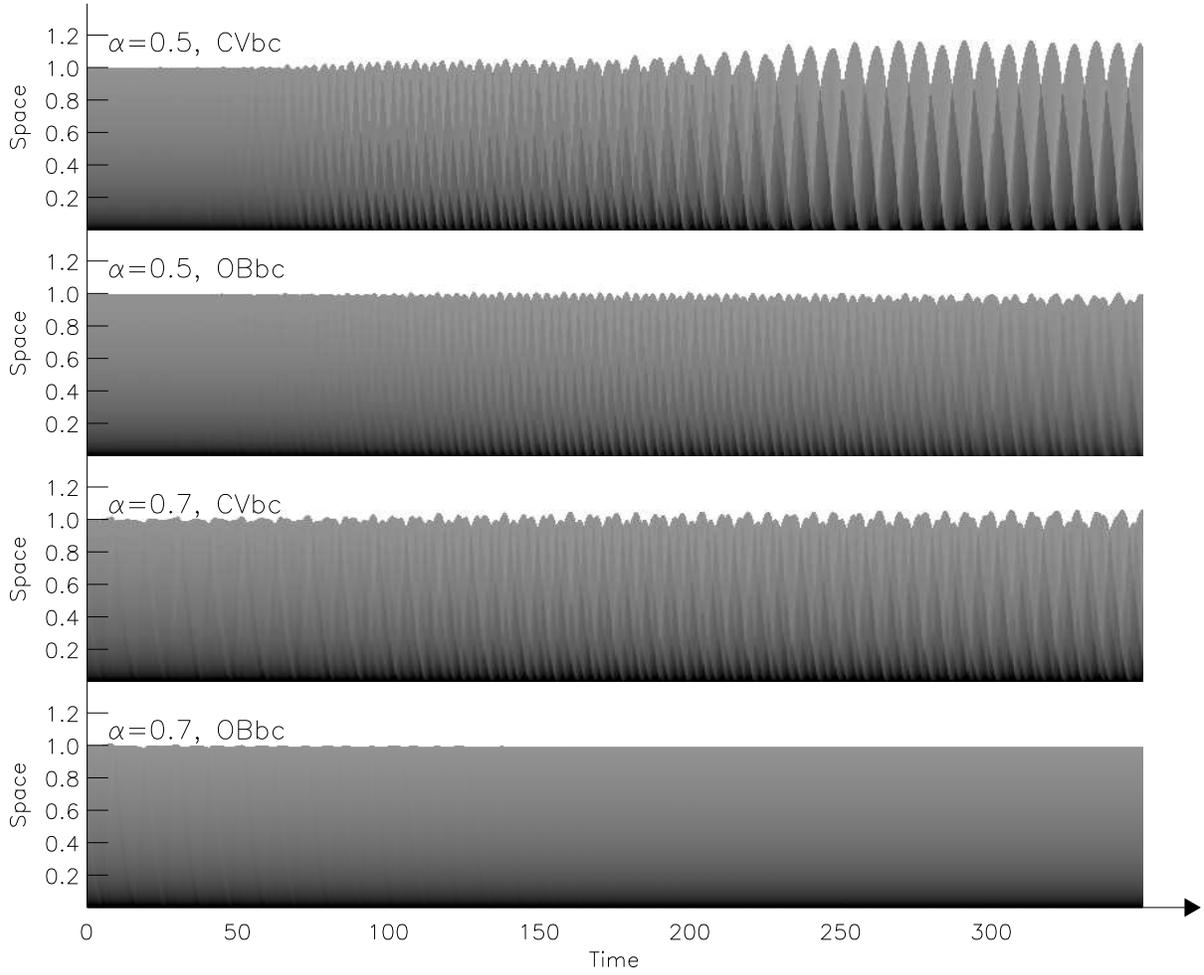}
\caption{\footnotesize Evolutionary space-time diagrams for 
         $\alpha = 0.5, 0.7$ with the OBbc and CVbc.
         Oscillation amplitudes are considerably reduced when 
         $\alpha$ increases and also when the
         OBbc is adopted. In the worst case (bottom panel),
         the initial perturbation is damped and the system returns
         to a stationary, stable configuration.
          \label{fig:alpha0507}}
\end{figure}

When the CVbc is adopted the system exhibits a linear phase 
for $t \lesssim 150$, gradually followed by the transition to 
the nonlinear regime.
When compared to the $\alpha=0$ case, the oscillation amplitudes in the
saturated regime are reduced by a factor of approximately $50\%$.
The situation is quite different, however, when the OBbc is considered:
Figure (\ref{fig:alpha0507}) shows that the solution remains close 
to the initial steady-state values and unstable oscillations grow
at a smaller rate.

A similar behavior has been reported by SB95 (who also adopted an 
``open boundary'' condition) for small values of the inflow Mach number 
(i.e., ${\cal M} = 5$). 
In their simulations, however, the amplitude of the oscillations 
was found to increase for higher Mach numbers, a behavior not 
observed in the present work.
The present conclusion is supported by several supplementary tests
in which both the inflow Mach number and the density of the cold
layer were changed, but a fully nonlinear growth of the instability
was still never observed.
In all the numerical tests, in fact, the cold gas layer always acts as an 
absorber to incoming perturbations, consequently reducing the amplitude
of the reflected waves.
Even in the presence of an external ``ad hoc'' perturbation 
(similar to the one introduced in SB95) it was found that the use of a 
cold dense ``layer'' inhibits the growth of instability when 
$\alpha \gtrsim 0.45$.

\begin{figure}
%\epsscale{.80}
\plotone{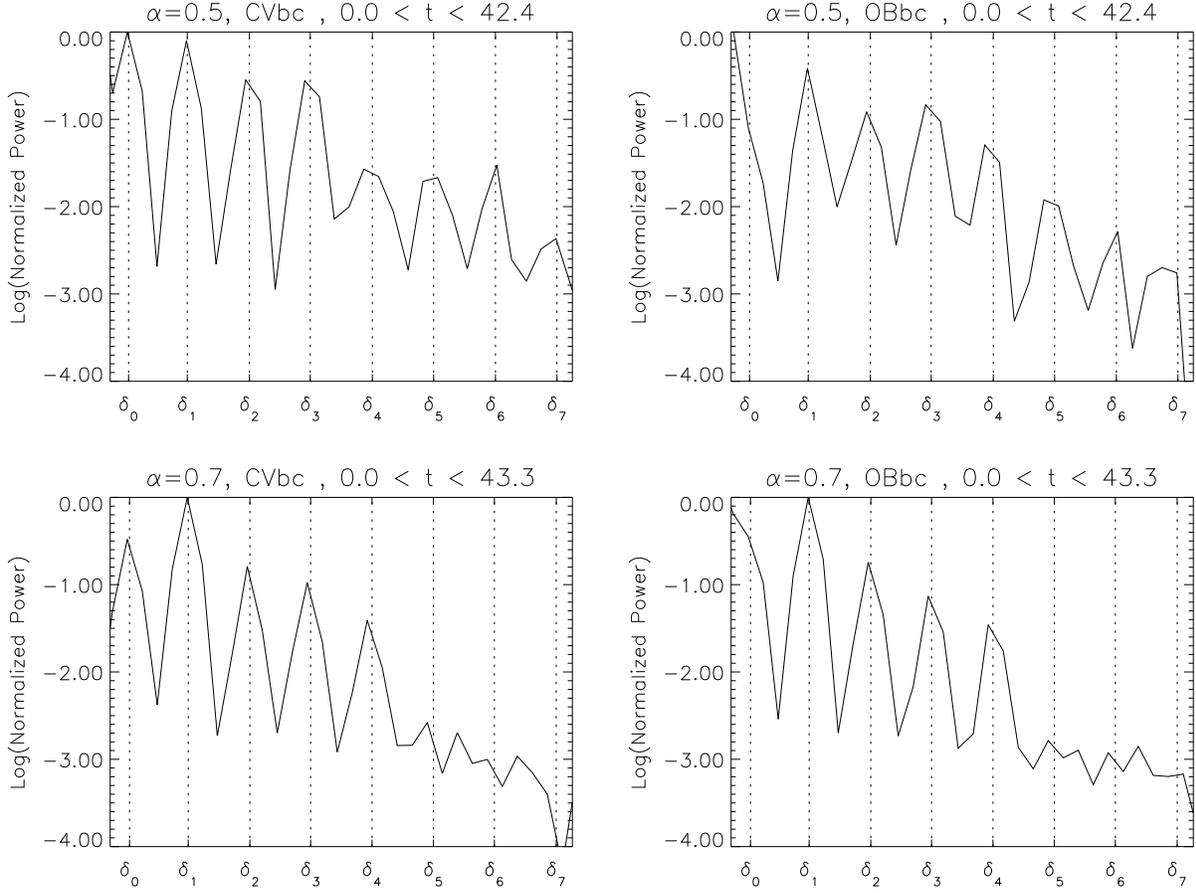}
\caption{\footnotesize Power spectra for $\alpha = 0.5$ (top) 
         and $\alpha = 0.7$ (bottom) during the early phases of 
         evolution, $0<t<42.4$ and $0<t<43.3$, respectively. 
         Results obtained with the constant velocity 
         and open boundary condition are shown on the left and on the right,
         respectively. The dashed vertical lines correspond to the
         frequencies derived from linear analysis. The vertical 
         and horizontal axis have the same meaning as in Fig. \ref{fig:alpha00_pow}.
         \label{fig:alpha0507_pow}}
\end{figure}

The behavior of the system during the early phases is reflected in the power
spectra shown in Figure \ref{fig:alpha0507_pow}, where a positive identification
of the oscillation eigenfrequencies with the linear ones is clear. 
The relative errors of the identifiable peaks are less than $4\%$ 
for all modes, see Table \ref{tab:error}.
Notice that the fundamental mode (expected to be stable from the
linear analysis) is also visible in the spectrum, since the initial 
numerical perturbation excites all modes regardless of their stability.

\begin{figure}
%\epsscale{.80}
\plotone{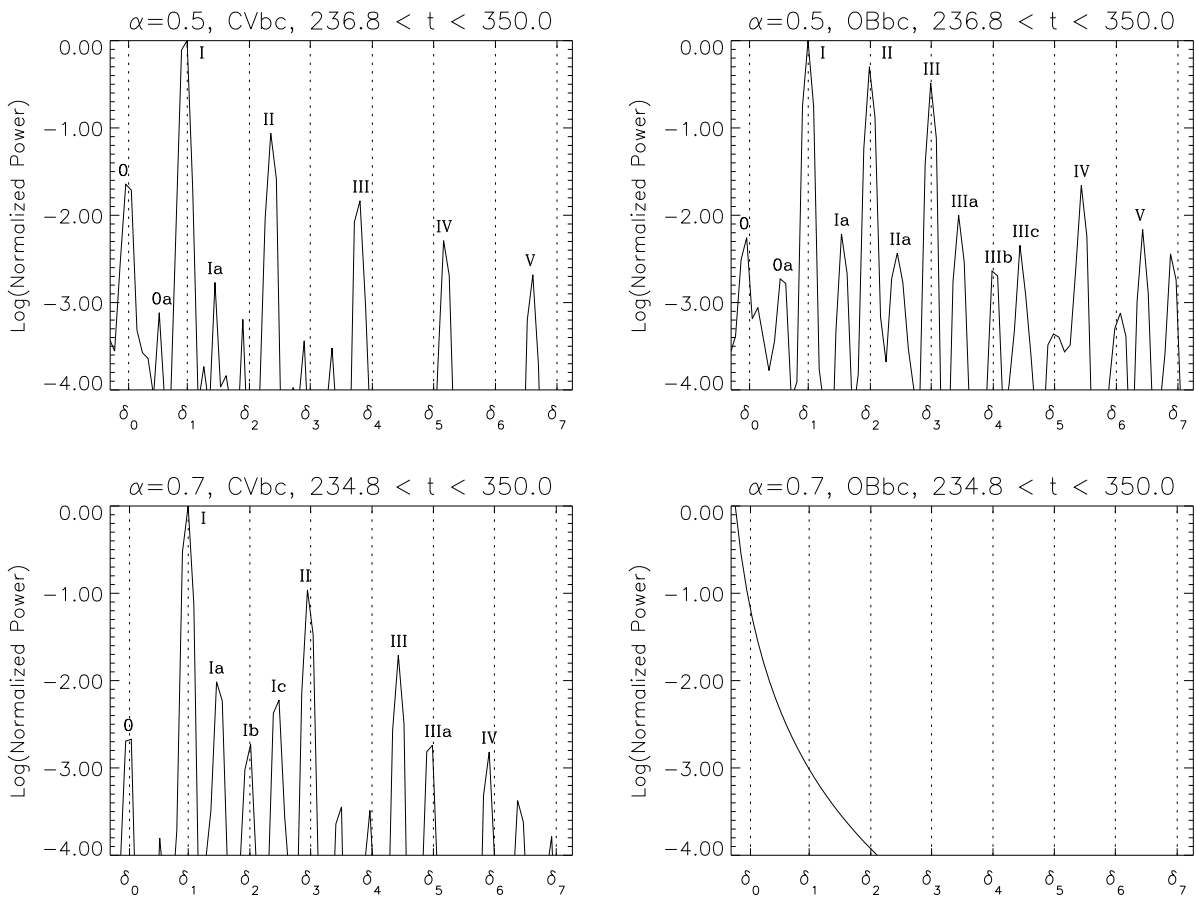}
\caption{\footnotesize Same as Fig. \ref{fig:alpha0507_pow}, but 
         for a longer portion of the oscillatory cycle during
         the late evolutionary phases, $236.8<t<350$ (for $\alpha = 0.5$)
         and $234.8<t<350$ (for $\alpha = 0.7$). During this
         time window, the oscillation amplitudes have fully saturated
         and a main sequence of modes, similar to those described in Fig. 
         \ref{fig:alpha00_pow_late}, may be distinguished
         in the power spectra. The corresponding oscillation
         frequencies are labeled by $\Omega^{\textrm{(I)}}$, $\Omega^{\textrm{(II)}}$, 
         $\Omega^{\textrm{(III)}}$, and so on.
         A number of secondary harmonics ($\Omega^\textrm{(Ia)}$, 
         $\Omega^{\textrm{(IIa)}}$, $\Omega^{\textrm{(IIb)}}$, etc.) is present as well. 
         Notice that the fundamental mode $\Omega^{(0)}$, although linearly
         stable, is still present (with little power) 
         in the spectra.
         Values of the different $\Omega$'s are listed
         in Table \ref{tab:nonlinear}.
         When $\alpha=0.7$ and the OBbc is used, the system returns to 
         steady-state and a flat spectrum is obtained.
         \label{fig:alpha0507_pow_late}}
\end{figure}

The power spectra taken during the later phases, Figure \ref{fig:alpha0507_pow_late},
reveal that, for the OBbc, the (mildly) unstable behavior is mostly
sustained by the first three overtones, whereas the first harmonic is
the only dominant mode for the CVbc.
In both cases, little contribution is given by the fundamental mode.
Nonlinear effects, however, suggest that the fundamental mode 
may still be important through mode-mode coupling.
Similarly to the $\alpha=0$ case, in fact, a main sequence of
modes can again be identified, see Figure \ref{fig:alpha0507_pow_late}.
For the CVbc, the frequency spacing between these modes is
either a multiple of the fundamental or equal to the first overtone,
e.g., $\Omega^{\textrm{(II)}} - \Omega^{\textrm{(I)}} \approx 3\Omega^{(0)}$ and
$\Omega^{\textrm{(IV)}} - \Omega^{\textrm{(III)}} \approx \Omega^{\textrm{(I)}}$. 
Secondary, small-power overtones are mainly visible for 
the OBbc case. Again, strong evidence for inter-mode coupling
is supported by the fact that these secondary overtones
may be decomposed into main sequence modes.
In fact, if one consider the frequencies listed in Table
\ref{tab:nonlinear}, it can be seen that
$\Omega^{\textrm{(0a)}} \approx \Omega^{\textrm{(I)}} - \Omega^{(0)}$,
$\Omega^{\textrm{(II)}} - \Omega^{\textrm{(Ia)}} \approx  \Omega^{(0)}$,
$\Omega^{\textrm{(IIIa)}} - \Omega^{\textrm{(III)}} \approx \Omega^{(0)}$,
and so on.

For $\alpha = 0.7$, an additional external perturbation has been introduced
to catalyze the onset of instability.
The perturbation is initially given in the velocity profile as
\begin{equation}\label{eq:perturbation}
   v_0(x) \quad  \Longrightarrow \quad
   v_0(x) \left(1 + \epsilon\exp\left[ -\left(\frac{x - 0.5}{0.1}\right)^2\right]\right)\,,
\end{equation}
with $\epsilon = 0.05$. Density and pressure are obtained
according to equations (\ref{eq:steady_1}) and (\ref{eq:steady_2}).

The different behaviors of the OBbc and CVbc are illustrated in
Figure \ref{fig:alpha0507}.
Results obtained with the OBbc show that the initial perturbation is damped 
roughly on a timescale $t \sim 150$. As for the $\alpha = 0.5$ case, the cold dense
layer behind the postshock region tends to quench large-amplitude perturbations. 
On the other hand, when the CVbc is adopted, the initial perturbation does
not fade away and the instability grows at a small rate. 
The amplitude of the oscillations relative to initial shock position is 
now further reduced to $\lesssim 5\%$ of the initial shock position.
Table \ref{tab:error} shows that the eigenfrequencies of the oscillations
differ by less than $8\%$ from the theoretical results.

The power spectra for the early linear phase, Figure
\ref{fig:alpha0507_pow}, are similar to the previous cases, 
although only modes with $0 \le n \le4$ (for the CVbc) and 
$1\le n\le 5$ (for the OBbc) contribute to the oscillations.

During the nonlinear phase (CVbc only), the first harmonic gives
the largest contribution, while the third overtone account for 
roughly $10\%$, Figure \ref{fig:alpha0507_pow_late}.
%-----------------------------------------------------------
Although little power is present in the fundamental mode, 
the frequency spacing between main sequence harmonics 
seems to indicate that mode-mode coupling may 
account, one more time, for the secondary, small-power intermediate 
peaks ($\Omega^{\textrm{(Ia)}}, \Omega^{\textrm{(Ib)}}$, etc.) shown in
Figure \ref{fig:alpha0507_pow_late}. 
Some the $\Omega$'s are, in fact, closely related:
$\Omega^{\textrm{(Ia)}} \approx \Omega^{\textrm{(II)}} - \Omega^{\textrm{(I)}}$, 
$\Omega^{\textrm{(Ic)}} \approx \Omega^{\textrm{(III)}} - \Omega^{\textrm{(Ia)}}$,
$\Omega^{\textrm{(IIIa)}} - \Omega^{\textrm{(III)}} \approx \Omega^{(0)}$,
$\Omega^{\textrm{(IV)}} - \Omega^{\textrm{(III)}} \approx  \Omega^{\textrm{(I)}}$,
and so on.
% ------------------------------------------------------------

%%%%%%%%%%%%%%%%%%%%%%%%%%%%%%%%%%%%%%%%%%%%%%%%%%%%%%%%%
\subsubsection{$\alpha = 0.8, 1$}
%
%
%%%%%%%%%%%%%%%%%%%%%%%%%%%%%%%%%%%%%%%%%%%%%%%%%%%%%%%%%

\begin{figure}
\plotone{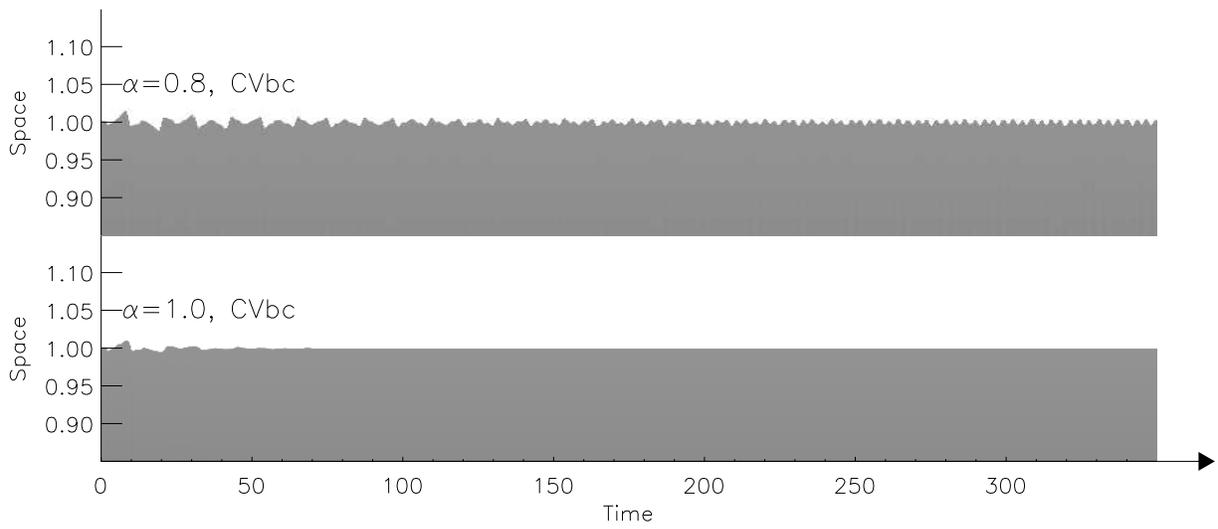}
\caption{\footnotesize Space-time evolutionary diagram for $\alpha = 0.8$ 
         (on the top) and $\alpha = 1$ (on the bottom) is shown.
         Only the CVbc has been employed. In order to make the oscillations
         more visible, the spatial scale in the plot shows a
         small area around the shock position. \label{fig:alpha081}}
\end{figure}

\begin{figure}
%\epsscale{.80}
\plotone{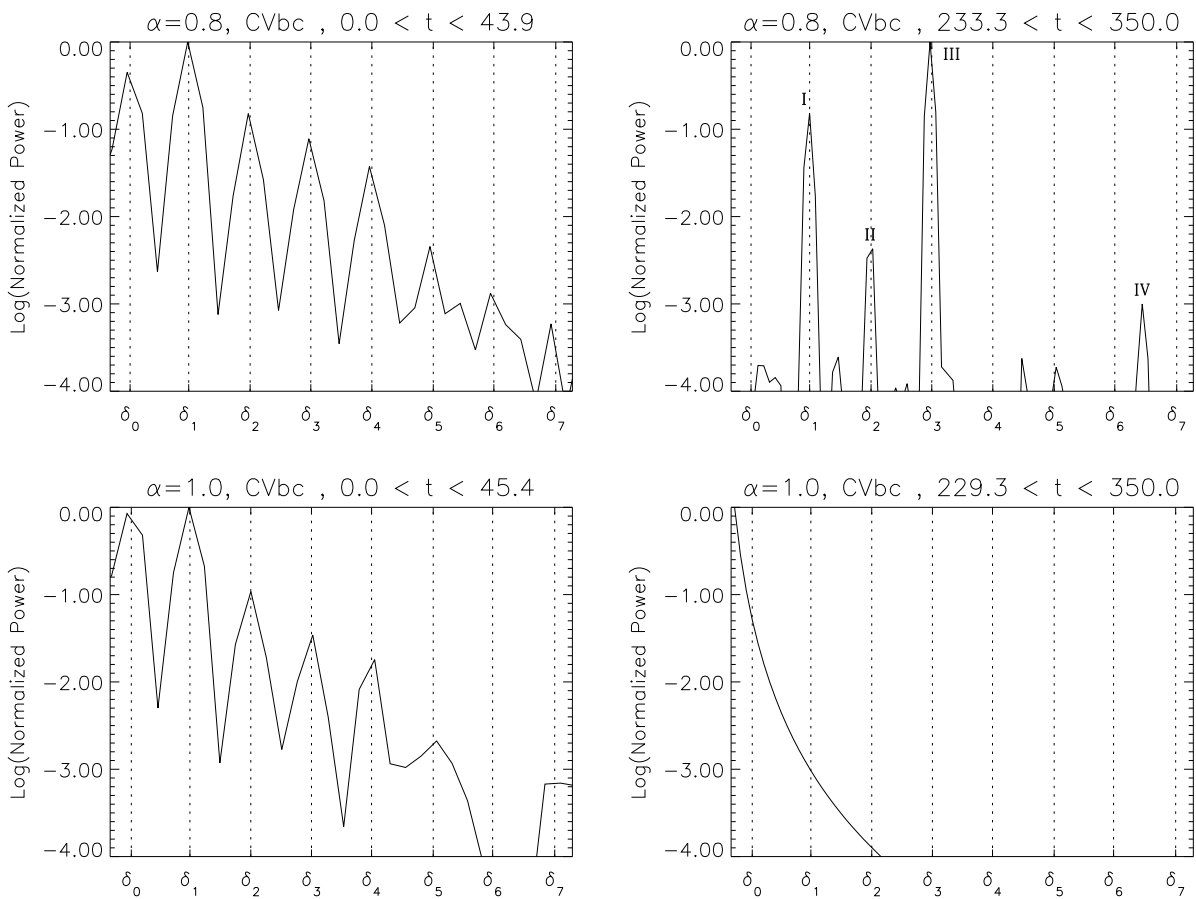}
\caption{\footnotesize Power spectra for $\alpha = 0.8$ (top panel) and 
         $\alpha = 1$ (bottom panel). The earlier ($0<t<43.9$ for $\alpha = 0.8$ and
         $0<t<45.4$ for $\alpha = 1$) and later ($233.3<t<350$ for $\alpha = 0.8$ and
         $229.3<t<350$ for $\alpha = 1$) simulation phases are shown to the left and
         to the right, respectively. 
         Only the $\alpha=0.8$ case evolves into a (weakly) nonlinear phase,
         characterized by very small-amplitude oscillations. 
         Notice how the first and second harmonics, although linearly 
         stable, are still present during the later simulation phases 
         ($\Omega^{\textrm{(I)}}$ and $\Omega^{\textrm{(II)}}$ on the top, right panel).
         The fundamental mode is practically absent.
         Explicit values of the nonlinear overtones are given in Table
         \ref{tab:nonlinear}. 
         When $\alpha = 1$, the initial perturbation produces modes
         that can be clearly identified with the linear ones (bottom, left panel).
         These modes are absent during the later phases (bottom, right panel), 
         thus confirming that the shock is stable. \label{fig:alpha081_pow}}
\end{figure}

The simulations for the last two cases, $\alpha=0.8$ and $\alpha = 1$,
are carried out using the CVbc only, since no growth of instability 
was observed using the OBbc.
In both cases, the perturbation given by equation (\ref{eq:perturbation}) 
was imposed at $t=0$. Results are shown in Figures 
\ref{fig:alpha081} and \ref{fig:alpha081_pow}.

When $\alpha=0.8$, the early phases of the evolution reflect the expected 
linear growth, as one can see from Figure \ref{fig:alpha081_pow}.
The power spectrum for to this phase shows modes of oscillations 
that clearly match the theoretical ones.
Most of the power is contributed by the first harmonic, followed by
the fundamental and then the remaining overtones.

The complete transition to the nonlinear phase occurs for $t \gtrsim 200$,
where smaller-amplitude, higher-frequency oscillations take over.
A power spectrum of the late evolution reveals the effects of this transition,
Figure \ref{fig:alpha081_pow}.
Most of the power is concentrated in the third harmonic, with only
$\sim 10 \% $ going into the first overtone and less than $\sim 1\%$ into
the second harmonic. The fundamental mode is absent. 

The fact that high-frequency oscillations are dominated by the third
harmonic is quite a remarkable result, since this overtone is the lowest 
unstable mode only in the narrow range $0.795 < \alpha < 0.85$,
while modes with $n < 3$ are stable as can be seen from Table \ref{tab:modes_2}.
This strongly suggests that this particular choice of boundary condition
is particularly consistent with linear results.

%------------------------------------------------------------
Notice that both the first and second harmonic should be 
linearly stable, since $\alpha_{\textrm c}^{(1)} = 0.7815 < 0.8$ and
$\alpha_{\textrm c}^{(2)} = 0.795 < 0.8$.
Their presence in the power spectrum, however, indicate
that a weak nonlinearity may probably be present.
Besides, nonlinear interactions are likely to be responsible
for the frequency coupling between the third and fourth main sequence
overtones, since $\Omega^{\textrm{(IV)}} \approx 2\Omega^{\textrm{(III)}}$.
%-----------------------------------------------------------------

Finally, when $\alpha = 1$, the initial perturbation is damped and the 
system returns to the original equilibrium solution for $t \gtrsim 70$. 
The power spectrum of the early evolutionary phase (Fig. \ref{fig:alpha081_pow})
shows that the shock oscillations are decomposed
into frequency modes that are well approximated by linear results.

%%%%%%%%%%%%%%%%%%%%%%%%%%%%%%%%%%%%%%%%%%%%%%%%%%%%%%%%%%%%%%%%%%%%%%
\section{DISCUSSION}
%
%
%
%
%
%%%%%%%%%%%%%%%%%%%%%%%%%%%%%%%%%%%%%%%%%%%%%%%%%%%%%%%%%%%%%%%%%%%%%%

A study of planar radiative shocks with a power-law cooling
function $\Lambda \sim \rho^2 T^\alpha$ has been conducted.
Both linear and nonlinear time-dependent calculations have 
been presented.

A linear stability analysis has been carried out for several 
values of the cooling index $\alpha$ in the range $[-2,2)$.
For a given value of $\alpha$, multiple discrete modes of oscillation exist 
and the real and imaginary parts of the first eight eigenfrequencies 
have been derived.
The overstable modes are labeled in order of increasing oscillation frequency
so that $n=0$ corresponds to the fundamental mode, $n=1$ to the first overtone,
$n=2$ to the second overtone, and so forth.
The stability criterion of a particular mode is expressed by the condition
$\alpha > \alpha^{(n)}_{\textrm c}$, where $\alpha^{(n)}_{\textrm c}$ is the critical value of the
cooling index for the $n$-th mode.
For the fundamental mode, for example, $\alpha^{(0)}_{\textrm c} = 0.388$, 
whereas for the first and second harmonic $\alpha^{(1)}_{\textrm c} = 0.782$ and 
$\alpha^{(2)}_{\textrm c} = 0.795$, respectively. 
A general trend towards stability exists for increasing $\alpha$,
so eventually all modes are stabilized for $\alpha \gtrsim 0.92$.
This study confirms previous results (CI), for which only the first 
two or three modes have been reported for some values of the cooling index.
It has been shown that oscillation frequencies are linearly proportional 
to the mode number $n$, a behavior similar to the quantized modes in 
a pipe.

The perturbative study has been complemented by several numerical simulations
using an Eulerian, high-resolution shock-capturing scheme.
The shock evolution has been followed through the linear and nonlinear
phases for different values of $\alpha$ and boundary conditions.
Among the four boundary conditions under consideration
(\S\ref{sec:boundary}), a new time-dependent boundary
treatment has been introduced. The new approach is based on the 
characteristic boundary method for the Euler equations and is particularly
consistent with the basic assumptions used in the analytical work, 
where the velocity perturbation has a node at the wall.

For the most unstable case considered here ($\alpha = 0$), all boundary
conditions yield similar results, although the reflective wall boundary 
condition is not particularly suitable in modeling small departures from
the stationary solution. 
The remaining three strategies provide modes of oscillations that,
in the limit of small perturbations, are close (within $5\%$ accuracy with
the exception of the fundamental mode) to the analytical values.

The cases $\alpha = 0.5, 0.7, 0.8$ and $1$ have also been considered.
For $\alpha=0.5,0.7$, the numerical simulations show that the choice of the 
lower boundary condition has a more severe impact on the growth of unstable 
modes during the saturated phase.
For example, the additional cold dense layer used in the open boundary condition
(\S\ref{sec:boundary}) tends to inhibit large-amplitude 
oscillations and, for $\alpha \gtrsim 0.7$, totally prevents the growth of
instability.  
In contrast, the new boundary approach yields results that closely 
reflect the analytical predictions; unstable behavior was observed for
$\alpha=0.5,0.7$, and $0.8$, although the saturated amplitude of
oscillations considerably decreases with increasing $\alpha$. 
For $\alpha = 0.8$, the largest oscillations during the saturated phase
are reduced to $\sim 0.5\%$ of the initial shock position and are thus 
barely visible at the resolution adopted (1600 zones for the postshock flow).
For $\alpha = 1$, the shock is stable and initial perturbations are damped 
on a characteristic timescale roughly proportional to the e-folding time
of the first overtone.
The modes of oscillations found in the numerical 
simulations (during the early phase of evolution) can be positively 
identified with the ones derived by linear analysis.
The relative error is usually small, $\lesssim 8\%$. Notice that this
error also accounts for the discrete frequency spacing 
introduced by the fast-Fourier transform of the shock position.

The transition from the linear to the nonlinear regime has also been 
investigated.
Power spectra of the shock position during the late evolutionary phases 
reveal that the first overtone is the dominant mode of oscillation when 
$\alpha \lesssim 0.7$, but that the third harmonic contributes to most
of the power at $\alpha = 0.8$.
The contribution of the fundamental is only $10\%$ for the most
unstable case ($\alpha = 0$), and decreases for increasing $\alpha$.

The new result of this work shows that a main sequence of
overtones characterizes the saturated, nonlinear oscillatory 
phases. Additional, secondary modes may also be
present, depending on the particular choice of boundary condition.
These modes of oscillation do not always match those predicted
by linear analysis but result from complex nonlinear interactions.
For the first time, evidence has been provided in favour of
mode-mode coupling, particularly between the first harmonic and the
fundamental mode.
The result extends also to those cases in which some of the modes 
are linearly stable ($\alpha = 0.5, 0.7,0.8$), thus supporting 
the possibility that linearly stable modes may actually 
become nonlinearly unstable. 

In summary, a general trend towards stability is found for $\alpha \gtrsim 0.8$, 
while an unstable behavior is expected for $\alpha \lesssim 0.4$,
regardless of the choice of the lower boundary condition.
On the other hand, numerical models of radiative shocks are more 
sensitive to the treatment of the lower boundary condition when 
$0.4 \lesssim \alpha \lesssim 0.8$, a range particularly relevant when 
optically thin bremsstrahlung is the dominant cooling mechanism.
It should be pointed out that the use of a cold layer of finite 
thickness may be more self-consistent in realistic astrophysical 
applications. The existence of the layer is, in fact, automatically
induced by a cutoff temperature in the cooling function and avoids
the complication of specifying a boundary conditions at the interface
between the postshock flow and the layer (provided
the sharp density gradients present in this region 
are adequately captured).

In spite of the oversimplifying assumptions adopted in this study,
these results show a number of interesting consequences for a 
variety of astrophysical settings.

Radiative shocks with velocities $v_{\textrm s} \gtrsim 130$ km s$^{-1}$ are
not uncommon in jets from young stellar objects, supernova remnants in 
the radiative phase, magnetospheric accretion in T-Tauri stars,
and colliding stellar winds in relatively close binary systems.
For these systems, the shocked interstellar gas reaches temperatures in the
range $10^5-10^7$ K and cools mainly by line emission, for which  
$\alpha < -0.5$.
Under these conditions, radiative shocks are likely to show
unstable behavior in all modes and phenomenological interpretations
based on steady-state models become of questionable validity
\citep{IGF87a, IGF87b}.
Although inclusion of transverse magnetic fields extends the range of 
stability \citep{Smith89, TD93}, the global thermal instability
of radiative shock waves may still be important in interpreting
a number of distinct observational features, such as emission-line ratios
observed in interstellar radiative shocks \citep{Hart94}, 
mixing between hot and cold material in colliding winds 
\citep{Stevens92,AOB04}, the filamentary structures observed in 
supernova remnants \citep{Blondin98, WF98}, and so forth.

Less conclusive assertions can be made for standing shocks in the 
accretion columns of Polar and Intermediate Polar systems.
At temperatures of the order of $10^8-10^9$ K the X-ray emission is primarily
determined by optically thin bremsstrahlung, although cyclotron and Compton
cooling may not be neglected \citep{Sax98}.
However, in the simple case where radiative losses are due to bremsstrahlung
cooling only, $\alpha \approx 0.5$, the dynamics of the shock may be influenced 
by the interaction with the upper photospheric layers of the white dwarf 
\citep{Crop90}.
Hence realistic models of accretion columns may require a more complex
treatment of the lower boundary.
For this reason, inclusion of additional physical processes 
such as magnetic fields, multi-dimensional effects, thermal conduction, etc., 
might be crucial for drawing firm conclusions about the stability 
of radiative shocks in AM-Her-type systems. Some of these issues 
will be considered in future extensions of this work.

\acknowledgements

The author would like to thank T. Plewa, B. Rosner, A. K{\" o}nigl,
D.~Q. Lamb, and T.J. Linde for their constructive support during  
my Ph.D. research program at the University of Chicago.

\appendix % ~~~~~~~~~~~~~~~~~~~~~~~~~~~~~~~~~~~~~~~~~~~~~~~~~~~~~~~~~~~~~~

%%%%%%%%%%%%%%%%%%%%%%%%%%%%%%%%%%%%%%%%%%%%%%%%%%%%%%%%%%%%%%%%%%%%%%%
\section{CHARACTERISTIC BOUNDARY CONDITIONS} \label{app:cvbc}
%
%
%
%
%
%%%%%%%%%%%%%%%%%%%%%%%%%%%%%%%%%%%%%%%%%%%%%%%%%%%%%%%%%%%%%%%%%%%%%%%

The hyperbolic nature of the Euler equations requires boundary 
conditions to be specified according to the way information propagates
in and out of the boundary.
The novel approach introduced in \S\ref{sec:boundary} is based on the
characteristic boundary method \citep{Thompson87,Thompson90},
where ``physical'' and ``numerical'' boundary conditions specify how 
zone boundary values are integrated in time along with the interior values. 
Although the subject of boundary conditions is a vast one and falls 
outside the scope of this paper, details of implementation are given
hereafter.

A ``physical'' boundary condition describes information that 
enters the domain and thus has to be imposed for each characteristic 
wave that propagates from the boundary towards the inside.
Information directed outside the boundary is entirely 
determined by the solution inside the domain and thus does not
require a boundary condition.
The numerical scheme, however, still depends on the knowledge of all 
flow variables at boundary zones, and hence additional ``numerical''
boundary conditions must be prescribed in a consistent way.

In the present context, the boundary equations are more conveniently
formulated using the quasi-linear form 
\begin{equation}\label{eq:bound_eqn}
 \pd{}{t}\left(\begin{array}{c}
  \tau  \\  \noalign{\medskip}
   v    \\  \noalign{\medskip}
   p    \\  \noalign{\medskip}
 \end{array}\right)  
    + 
 \left(\begin{array}{ccc} 
  v  & -\tau    & 0     \\  \noalign{\medskip}
  0  &   v      & \tau  \\  \noalign{\medskip}
  0  & \gamma p &  v    \\
  \end{array}\right) \pd{}{x}\left(\begin{array}{c}
  \tau  \\  \noalign{\medskip}
   v    \\  \noalign{\medskip}
   p    \\  \noalign{\medskip}
 \end{array}\right) = \left(\begin{array}{c} 
    0   \\  \noalign{\medskip}
    0   \\  \noalign{\medskip}
  -(\gamma - 1)\Lambda \end{array}\right) 
\end{equation}
where $\tau = 1/\rho$. The system of equations (\ref{eq:bound_eqn})
holds at the boundary and must evolve in time together with the
interior values. 
The characteristic speeds of the system (\ref{eq:bound_eqn}) are given by
$\lambda^\pm = v \pm c_s$, $\lambda^0 = v$, where $c_s = (\gamma p /\rho)^{1/2}$
is the speed of sound.

In the postshock region the flow is initially subsonic everywhere, 
since $-c_s < v < 0$ for $0 < x < 1$.
In the limit of small perturbations around the steady-state values, 
it is reasonable to assume that a condition for subsonic outflow will
continue to hold at subsequent times.
Hence, the characteristic associated with $\lambda^+$ has positive sign
(i.e., it carries information inside the domain), whereas $\lambda^0$ and
$\lambda^-$ are directed outward.
This means that only one ``physical'' boundary condition can be freely 
specified (e.g., a constant pressure or velocity) and the remaining two must
be compatible with the interior discretization scheme.
Choosing a constant outflow velocity, for example, is consistent with 
the linear perturbative analysis, where the velocity perturbation is
forced to have a node at the origin.

Integration of the boundary equations (\ref{eq:bound_eqn}) proceeds
by splitting the time-dependent solution into a contribution
coming from the steady-state value and a time-dependent 
``deviation'': 
\begin{equation} 
 q(x,t) = q_1(x,t) + q_0(x)
\end{equation}
where $q\in \left\{\tau, u, p\right\}$.
In the ``constant-velocity'' boundary condition, the velocity 
perturbation $v_1 = 0$ at all times and the boundary equations 
prescribe how pressure and density should evolve with time:
\begin{equation}\label{eq:bound_eq_1}
 \pd{\tau_1}{t} + v_0\pd{\tau_1}{x} - \tau\pd{v_1}{x} = \tau_1\pd{v_0}{x} \,,
\end{equation}
\begin{equation}\label{eq:bound_eq_2}
 \pd{p_1}{t} + v_0\pd{p_1}{x} + \gamma p\pd{v_1}{x} = 
  -(\gamma - 1)\left(\Lambda - \Lambda_0\right) - \gamma p_1\pd{v_0}{x} \,,
\end{equation}
where the spatial derivatives are computed using one-sided
approximations.
Notice that equations (\ref{eq:bound_eq_1}) and (\ref{eq:bound_eq_2}) 
are not a linearization around a stationary solution but are, 
in principle, valid for arbitrary departures.

%%%%%%%%%%%%%%%%%%%%%%%%%%%%%%%%%%%%%%%%%%%%%%%%%%%%%%%%%%%%%%%%%%%%%%%%%%%%%
% 
%                        R E F E R E N C E S
%
%%%%%%%%%%%%%%%%%%%%%%%%%%%%%%%%%%%%%%%%%%%%%%%%%%%%%%%%%%%%%%%%%%%%%%%%%%%%%

%%%%%%%%%%%%%%%%%%%%%%%%%%%%%%%%%%%%%%%%%%%%%%%%%%%%%%%%%%%%%%%%%%%%%%%%%
% 
%                    F I G U R E S
%
%%%%%%%%%%%%%%%%%%%%%%%%%%%%%%%%%%%%%%%%%%%%%%%%%%%%%%%%%%%%%%%%%%%%%%%%%

% --------------------
%   Linear analysis
% --------------------

\clearpage

\clearpage

\clearpage

\clearpage

% --------------------
%  Numerical results
% --------------------

\clearpage

\clearpage

\clearpage

% ---- \alpha = 0.5, 0.7 ---- 

\clearpage

\clearpage

\clearpage

% ---- \alpha = 0.8, 1.0 ---- 

\clearpage

\clearpage

\input{tab1.tex}

\end{document}

%% file: tab1.tex
%%%%%%%%%%%%%%%%%%%%%%%%%%%%%%%%%%%%%%%%%%%%%%%%%%%%%%%%%%%%%%%%%%%%%%%%%%%%%
% 
%                              T A B L E S   
%
%%%%%%%%%%%%%%%%%%%%%%%%%%%%%%%%%%%%%%%%%%%%%%%%%%%%%%%%%%%%%%%%%%%%%%%%%%%%%

\clearpage
%%%%%%%%%%%%%%%%%%%%%%%%%%%%%%%%%%%%%%%%%%%%%%%%%%%%%%%%%%%%%%%%
%
%   Real and imaginary parts (1)
%
%                   000000000111
%                   123456789012
\begin{deluxetable}{cccccccccccc}
\tabletypesize{\footnotesize}
\tablecolumns{12} 
\tablewidth{0pt} 
\tablecaption{Real and imaginary parts of the complex eigenfrequencies 
              $\delta = \delta_{\textrm R} + i\delta_{\textrm I}$ are
              given for the first eight modes, $n=0..7$, and for negative
              values of $\alpha$. The lower portion of the Table shows 
              the coefficients derived from the linear fit 
              $\delta_{\textrm I}^{(n)} = \td^{(0)} + n\Delta\td$, where
              $\td^{(0)}$ is the ``fitted'' fundamental mode and 
              $\Delta\td$ is the frequency spacing. \label{tab:modes_1}}
\tablehead{
\colhead{} & \multicolumn{2}{c}{$\alpha = -2$}   &   \colhead{}   &
             \multicolumn{2}{c}{$\alpha = -3/2$} &   \colhead{}   &
             \multicolumn{2}{c}{$\alpha = -1$}   &   \colhead{}   &
             \multicolumn{2}{c}{$\alpha = -1/2$} \\
\cline{2-3} \cline{5-6} \cline{8-9} \cline{11-12}\\
 \colhead{Mode} & \colhead{$\delta_{\textrm R}$} & \colhead{$\delta_{\textrm I}$} &
 \colhead{}     & \colhead{$\delta_{\textrm R}$} & \colhead{$\delta_{\textrm I}$} & 
 \colhead{}     & \colhead{$\delta_{\textrm R}$} & \colhead{$\delta_{\textrm I}$} & 
 \colhead{}     & \colhead{$\delta_{\textrm R}$} & \colhead{$\delta_{\textrm I}$}} 
\startdata
$n = 0$ & 0.1671 & 0.2175 & & 0.1353 & 0.2416 & & 0.1031 & 0.2616 & & 0.0693 & 0.2787 \\ 
$n = 1$ & 0.3443 & 0.9581 & & 0.2925 & 0.9566 & & 0.2393 & 0.9510 & & 0.1827 & 0.9399  \\ 
$n = 2$ & 0.3905 & 1.7252 & & 0.3360 & 1.7052 & & 0.2786 & 1.6778 & & 0.2161 & 1.6398  \\ 
$n = 3$ & 0.4258 & 2.4622 & & 0.3707 & 2.4277 & & 0.3121 & 2.3820 & & 0.2476 & 2.3204  \\ 
$n = 4$ & 0.4538 & 3.2200 & & 0.3957 & 3.1704 & & 0.3334 & 3.1059 & & 0.2642 & 3.0197 \\ 
$n = 5$ & 0.4684 & 3.9594 & & 0.4110 & 3.8945 & & 0.3495 & 3.8112 & & 0.2812 & 3.7012 \\ 
$n = 6$ & 0.4918 & 4.7110 & & 0.4319 & 4.6330 & & 0.3669 & 4.5325 & & 0.2938 & 4.3996 \\ 
$n = 7$ & 0.4982 & 5.4548 & & 0.4389 & 5.3604 & & 0.3749 & 5.2400 & & 0.3039 & 5.0819 \\ 
\tableline\tableline \\[1pt]
        & $\td^{(0)}$ & $\Delta\td$ & & $\td^{(0)}$ & $\Delta\td$ & & $\td^{(0)}$ & $\Delta\td$ & & $\td^{(0)}$ & $\Delta\td$ \\[3pt]   
\tableline  \\
       & 0.2183 & 0.7486 & & 0.2352 & 0.7324 & & 0.2502 & 0.7129 & & 0.2642 & 0.6882  \\
\enddata

%% Text for table notes should follow after the \enddata but before
%% the \end{deluxetable}. Make sure there is at least one \tablenotemark
%% in the table for each \tablenotetext.
% \tablecomments{}
\end{deluxetable}

%%%%%%%%%%%%%%%%%%%%%%%%%%%%%%%%%%%%%%%%%%%%%%%%%%%%%%%%%%%%%%%%
%
%   Real and imaginary parts (2)
%
%
%                   0000000001111
%                   1234567890123
\begin{deluxetable}{ccccccccccccc}
\tabletypesize{\footnotesize}
\tablecolumns{12} 
\tablewidth{0pt} 
\tablecaption{Real and imaginary parts of the complex eigenfrequencies 
              $\delta = \delta_{\textrm R} + i\delta_{\textrm I}$ are
              given for the first eight modes, $n=0..7$, and for nonnegative
              values of $\alpha$. The rightmost column gives the critical 
              value of $\alpha$ for a given mode $n$, such that for 
              $\alpha > \alpha_{\textrm c}^{(n)}$ the $n$-th mode is stable.
              The lower portion of the Table lists the coefficients derived
              from the linear fit. \label{tab:modes_2}}
\tablehead{
\colhead{} & \multicolumn{2}{c}{$\alpha = 0$}   &   \colhead{}   &
             \multicolumn{2}{c}{$\alpha = 1/2$} &   \colhead{}   &
             \multicolumn{2}{c}{$\alpha = 1$}   &   \colhead{}   &
             \multicolumn{2}{c}{$\alpha = 3/2$} &    \colhead{$\alpha_{\textrm c}$}  \\
\cline{2-3} \cline{5-6} \cline{8-9} \cline{11-12}\\
 \colhead{Mode} & \colhead{$\delta_{\textrm R}$} & \colhead{$\delta_{\textrm I}$} &
 \colhead{}     & \colhead{$\delta_{\textrm R}$} & \colhead{$\delta_{\textrm I}$} & 
 \colhead{}     & \colhead{$\delta_{\textrm R}$} & \colhead{$\delta_{\textrm I}$} & 
 \colhead{}     & \colhead{$\delta_{\textrm R}$} & \colhead{$\delta_{\textrm I}$} & }
\startdata
$n = 0$ & 0.0323  & 0.2934 & & -0.0101  & 0.3052 & & -0.0622  & 0.3121 & & -0.1346  & 0.3054 & 0.3881 \\ 
$n = 1$ & 0.1201  & 0.9210 & & 0.0476  & 0.8887 & & -0.0420  & 0.8307 & & -0.1668  & 0.7075 & 0.7815 \\ 
$n = 2$ & 0.1450  & 1.5857 & & 0.0602  & 1.5043 & & -0.0485  & 1.3698 & & -0.2020  & 1.1022 & 0.7949 \\ 
$n = 3$ & 0.1739  & 2.2347 & & 0.0851  & 2.1087 & & -0.0310  & 1.9068 & & -0.2141  & 1.5186 & 0.8822 \\ 
$n = 4$ & 0.1841  & 2.9003 & & 0.0865  & 2.7242 & & -0.0396  & 2.4386 & & -0.2128  & 1.9042 & 0.8578 \\ 
$n = 5$ & 0.2021  & 3.5505 & & 0.1052  & 3.3322 & & -0.0280  & 2.9850 & & -0.2416  & 2.3021 & 0.9111 \\ 
$n = 6$ & 0.2081  & 4.2158 & & 0.1030  & 3.9454 & & -0.0307  & 3.5106 & & -0.2390  & 2.7186 & 0.8959 \\ 
$n = 7$ & 0.2214  & 4.8666 & & 0.1188  & 4.5561 & & -0.0286  & 4.0605 & & -0.2375  & 3.0974 & 0.9196 \\
\tableline\tableline \\[1pt]
  & $\td^{(0)}$& $\Delta\td$ & & $\td^{(0)}$& $\Delta\td$ & & $\td^{(0)}$& $\Delta\td$ & & $\td^{(0)}$& $\Delta\td$ & \\[3pt] 
\tableline  \\
     & 0.2774 & 0.6553 & & 0.2898 & 0.6088 & & 0.3011 & 0.5359 & & 0.3076 & 0.3998 &   \\
\enddata

%% Text for table notes should follow after the \enddata but before
%% the \end{deluxetable}. Make sure there is at least one \tablenotemark
%% in the table for each \tablenotetext.
\end{deluxetable}

%%%%%%%%%%%%%%%%%%%%%%%%%%%%%%%%%%%%%%%%%%%%%%%%%%%%%%%%%%%%%%%%%%%
%
%    RELATIVE ERRORS DURING LINEAR PHASE
%
%
%                   0000000000123
%                   1234567891111
\begin{deluxetable}{ccccccccccccc}
\tabletypesize{\footnotesize}
\tablewidth{0pt}
\tablecaption{Relative errors of the oscillation frequencies found from 
              the numerical simulations during the early linear phases.
              The errors are computed as $|\omega^{(n)}_{\textrm I}/\delta^{(n)} - 1|$,
              where $\omega^{(n)}_{\textrm I}$ 
              corresponds to the closest frequency peak referred to 
              the theoretical value.
              Notice that the finite length of the time window $\Delta t$ over
              which the Fourier transform is taken introduces an 
              uncertainty $\sim 1/\Delta t$.\label{tab:error}}
\tablehead{
\colhead{} & \multicolumn{2}{c}{$\alpha = 0$}   &   \colhead{}   &
             \multicolumn{2}{c}{$\alpha = 1/2$} &   \colhead{}   &
             \multicolumn{2}{c}{$\alpha = 0.7$} &   \colhead{}   &
             \multicolumn{1}{c}{$\alpha = 0.8$} &   \colhead{}   &
             \multicolumn{1}{c}{$\alpha = 1$}  \\
\cline{2-3}  \cline{5-6}  \cline{8-9} \cline{11-11} \cline{13-13} \\
 \colhead{Mode} & \colhead{CVbc} & \colhead{OBbc} &
 \colhead{}     & \colhead{CVbc} & \colhead{OBbc} & 
 \colhead{}     & \colhead{CVbc} & \colhead{OBbc} &
 \colhead{}     & \colhead{CVbc} & 
  \colhead{}    & \colhead{CVbc}} 
\startdata
$n = 0$ & 0.482 & 0.036 & & 0.039 &    -  & & 0.071 &    -  & & 0.087 & & 0.122 \\ 
$n = 1$ & 0.010 & 0.010 & & 0.008 & 0.008 & & 0.010 & 0.010 & & 0.010 & & 0.010 \\
$n = 2$ & 0.042 & 0.042 & & 0.025 & 0.025 & & 0.017 & 0.017 & & 0.012 & & 0.0003 \\
$n = 3$ & 0.048 & 0.020 & & 0.026 & 0.026 & & 0.016 & 0.016 & & 0.009 & & 0.006 \\
$n = 4$ & 0.005 & 0.005 & & 0.031 & 0.031 & & 0.017 & 0.017 & & 0.009 & & 0.011 \\
$n = 5$ & 0.016 & 0.027 & & 0.012 & 0.032 & & 0.018 & 0.018 & & 0.010 & & 0.010 \\
$n = 6$ & 0.009 & 0.009 & & 0.004 & 0.004 & & 0.018 & 0.018 & & 0.008 & &   -   \\
$n = 7$ & 0.032 & 0.001 & & 0.035 & 0.035 & & 0.046 & 0.013 & & 0.010 & & 0.012 \\
\enddata

%% Text for table notes should follow after the \enddata but before
%% the \end{deluxetable}. Make sure there is at least one \tablenotemark
%% in the table for each \tablenotetext.
% \tablecomments{}
\end{deluxetable}

%%%%%%%%%%%%%%%%%%%%%%%%%%%%%%%%%%%%%%%%%%%%%%%%%%%%%%%%%%%%%%%%%%%
%
%    NONLINEAR FREQ.
%
%
%                   0000000000
%                   1234567891
\begin{deluxetable}{cccccccccc}
\tabletypesize{\footnotesize}
\tablewidth{0pt}
\tablecaption{ Nonlinear frequencies relative to the later
              evolutionary phases for $\alpha=0, 0.5, 0.7$ and $0.8$.
              Frequencies labeled with $\Omega^\textrm{(I)}$, $\Omega^\textrm{(II)}$,
              $\Omega^\textrm{(III)}$, and so on, identify the main sequence 
              overtones, whereas the intermediate secondary modes 
              are enumerated by appending a letter to the main sequence 
              mode number that precedes it (i.e. $\Omega^\textrm{(Ia)}$, $\Omega^\textrm{(IIa)}$,
              $\Omega^\textrm{(IIb)}$, etc.)
              The error introduced by the Fourier transform is 
              $\sim 1/\Delta t$, where $\Delta t$ is the time 
              window over which the transform is taken.\label{tab:nonlinear}}
\tablehead{
\colhead{} & \multicolumn{2}{c}{$\alpha = 0$}   &   \colhead{}   &
             \multicolumn{2}{c}{$\alpha = 1/2$} &   \colhead{}   &
             \multicolumn{1}{c}{$\alpha = 0.7$} &   \colhead{}   &
             \multicolumn{1}{c}{$\alpha = 0.8$} \\
\cline{2-3}  \cline{5-6}  \cline{8-8} \cline{10-10} \\
 \colhead{Mode} & \colhead{CVbc} & \colhead{OBbc} &
 \colhead{}     & \colhead{CVbc} & \colhead{OBbc} & 
 \colhead{}     & \colhead{CVbc} & 
 \colhead{}     & \colhead{CVbc} } 
\startdata
$\Omega^{(0)}$           & 0.287 & 0.287 & & 0.278 & 0.278 & & 0.326 &  &   -    \\
$\Omega^\textrm{(0a)}$   & 0.575 &   -   & & 0.611 & 0.611 & &   -   &  &   -    \\
$\Omega^\textrm{(I)}$    & 0.805 & 0.920 & & 0.888 & 0.888 & & 0.869 &  & 0.858  \\
$\Omega^\textrm{(Ia)}$   & 1.092 &   -   & & 1.166 & 1.222 & & 1.141 &  &   -    \\
$\Omega^\textrm{(Ib)}$   & 1.379 &   -   & & 1.444 &   -   & & 1.467 &  &   -    \\
$\Omega^\textrm{(Ic)}$   &   -   &   -   & &   -   &   -   & & 1.738 &  &   -    \\
$\Omega^\textrm{(II)}$   & 1.667 & 1.839 & & 1.721 & 1.499 & & 2.010 &  & 1.448  \\
$\Omega^\textrm{(IIa)}$  & 1.897 &   -   & &   -   & 1.777 & &   -   &  &   -    \\
$\Omega^\textrm{(IIb)}$  & 2.184 &   -   & &   -   &   -   & &   -   &  &   -    \\
$\Omega^\textrm{(III)}$  & 2.471 & 2.759 & & 2.554 & 2.110 & & 2.879 &  & 1.985  \\
$\Omega^\textrm{(IIIa)}$ & 2.759 &   -   & &   -   & 2.388 & & 3.205 &  &   -    \\
$\Omega^\textrm{(IIIb)}$ & 3.046 &   -   & &   -   & 2.721 & &   -   &  &   -    \\
$\Omega^\textrm{(IIIc)}$ &   -   &   -   & &   -   & 2.999 & &   -   &  &   -    \\
$\Omega^\textrm{(IV)}$   & 3.276 & 3.678 & & 3.443 & 3.609 & & 3.748 &  & 3.969  \\
$\Omega^\textrm{(V)}$    & 4.138 & 4.598 & & 4.331 & 4.220 & &   -   &  &   -    \\
\enddata

%% Text for table notes should follow after the \enddata but before
%% the \end{deluxetable}. Make sure there is at least one \tablenotemark
%% in the table for each \tablenotetext.
% \tablecomments{}
\end{deluxetable}

%% file: ms.bbl
\begin{thebibliography}{}

\bibitem[Antokhin et al.(2004)]{AOB04}
    Antokhin, I.~I., Owocki, S.~P., \& Brown, J.~C.\
    2004, \apj, 611, 434

\bibitem[Bertschinger(1986)]{Bert86} 
     Bertschinger, E.\
     1986, \apj, 304, 154

\bibitem[Blondin et al.(1998)]{Blondin98}
     Blondin, J.~M., Wright, E.~B., Borkowski, K.~J., \& Reynolds, S.~P.\ 
     1998, \apj, 500, 342 

\bibitem[Calvet \& Gullbring(1998)]{CG98}
     Calvet, N.~\& Gullbring, E.\ 
     1998, \apj, 509, 802 

\bibitem[Chanmugam et al.(1985)]{Chan85} 
    Chanmugam, G., Langer, S.~H., \& Shaviv G.\
    1985, \apj, 299, L87

\bibitem[Chevalier \& Imamura(1982)]{CI82} 
    Chevalier, R.~A.~\& Imamura, J.~M.\
     2003, \apj, 261, 543

\bibitem[Colella(1990)]{Colella90}
     Colella, P.\
     1990,\apj, 54,174.

\bibitem[Cropper(1990)]{Crop90}
     Cropper, M.\ 
     1990, \ssr, 54, 195
 
\bibitem[Dgani \& Soker(1994)]{DS94}
     Dgani, R., \& Soker, N.\ 
     1994, \apj, 434, 262 

\bibitem[Edelman(1989a)]{Ed89a}
     Edelman, M.~A.\
     1989, Astrofizika, 31, 407 (English transl., Astrophysics,31 656 [1990])

\bibitem[Edelman(1989b)]{Ed89b}
     Edelman, M.~A.\
     1989, Astrofizika, 31, 579 (English transl., Astrophysics,31 758 [1990])

\bibitem[Gaetz et al.(1988)]{GEC88} 
    Gaetz, T.~J., Edgar, R.~J., \& Chevalier, R.~A.\
    1988, \apj, 329, 927

\bibitem[Gottlieb \& Shu(1998)]{Gottlieb_Shu98}
     Gottlieb, S.~\& Shu, C.~W.\ 
     1998, Math. Comput., 67,  73

\bibitem[Hartigan et al.(1994)]{Hart94}
     Hartigan, P., Morse, J.~A., \& Raymond, J.\
     1994, \apj, 436, 125

\bibitem[Houck \& Chevalier(1992)]{HC92}
     Houck, J.~C.~\& Chevalier, R.~A.\
     1992, \apj, 395, 592

\bibitem[Hujeirat \& Papaloizou(1998)]{HP98}
     Hujeirat, A., \& Papaloizou, J.~C.~B.\ 
     1998, \aap, 340, 593 

\bibitem[Imamura et al.(1984)]{Ima84} 
     Imamura, J.~N., Wolff, M.~T., \& Durisen, R.~H.\
     1984, \apj, 276, 667

\bibitem[Imamura(1985)]{Ima85} 
     Imamura, J.~N.\
     1985, \apj, 296, 128

\bibitem[Imamura et al.(1991)]{Ima91}
     Imamura, J.~N., Rashed, H., \& Wolff, M.~T.\ 
     1991, \apj, 378, 665 

\bibitem[Imamura et al.(1996)]{Ima96} 
     Imamura, J.~N., Aboasha, A., Wolff, M.~T. \& Wood, S.~W.\
     1996, \apj, 458, 327

\bibitem[Innes et al.(1987a)]{IGF87a} 
    Innes, D.~E., Gidding, J.~R., \& Falle, S.~A.~E.~G.\
    1987, \mnras, 226, 67

\bibitem[Innes et al.(1987b)]{IGF87b} 
    Innes, D.~E., Gidding, J.~R., \& Falle, S.~A.~E.~G.\
    1987, \mnras, 227, 1021

\bibitem[Kimoto \& Chernoff(1997)]{KC97}
    Kimoto, P.~A., \& Chernoff, D.~F.\ 
    1997, \apj, 485, 274 

\bibitem[Langer et al.(1981)]{LCS81} 
    Langer, S.~H., Chanmugam, G., \& Shaviv, G.\ 
    1981, \apj, 245, L23

\bibitem[Langer et al.(1982)]{LCS82} 
    Langer, S.~H., Chanmugam, G., \& Shaviv, G.\
    1982, \apj, 258, 289

\bibitem[Langer et al.(1983)]{LCS83} 
    Langer, S.~H., Chanmugam, G., \& Shaviv, G.\
    1982, in Cataclysmic Variables and Related Objects,
    ed. M. Livio and G. Shaviv (Dordrecht: Reidel), p. 199

\bibitem[Larsson(1992)]{Larsson92}
     Larsson, S.\
     1992, \aap, 265, 133 

\bibitem[LeVeque(1998)]{LeVeque98}
     LeVeque, R.~J., Mihalas, D., Dorfi, E.~A., \& M\"uller, E.\
     1998, Computational Methods for Astrophysical Flow:
     Springer-Verlag

\bibitem[Middleditch et al.(1997)]{Middleditch97}
        Middleditch, J., Imamura, J.~N., \& Steiman-Cameron, T.~Y.\
        1997, \apj, 489, 912 

\bibitem[Plewa(1995)]{Plewa95}
         Plewa, T.\ 
         1995, \mnras, 275, 143
\bibitem[Roe(1981)]{Roe81}
     Roe, P.~L.\
     1981, J. Comp. Phys., 43, 357

\bibitem[Saxton et al.(1997)]{Sax97}
     Saxton, C.~J., Wu, K., \& Pongracic, H.\
     1997, \pasa, 14, 164 

\bibitem[Saxton et al.(1998)]{Sax98}
     Saxton, C.~J., Wu, K., Pongracic, H., \& Shaviv, G.\
     1998, \mnras, 299, 862 

\bibitem[Saxton(1999)]{Sax99}
     Saxton, C.~J.\
     1999, Ph.d. Thesis, University of Sydney.

\bibitem[Saxton \& Wu(1999)]{SaxWu99}
     Saxton, C.~J., \& Wu, K.\
     1999, \mnras, 310, 677 

\bibitem[Saxton \& Wu(2001)]{SaxWu01}
     Saxton, C.~J., \& Wu, K.\
     2001, \mnras, 324, 659 

\bibitem[Saxton(2002)]{Sax02}
     Saxton, C.~J.\
     2002, \pasa, 19, 282 

\bibitem[Smith(1989)]{Smith89}
     Smith, M.~D.\
     1989, \mnras, 238, 235 

\bibitem[Stevens et al.(1992)]{Stevens92}
     Stevens, I.~R., Blondin, J.~M., \& Pollock, A.~M.~T.\ 
     1992, \apj, 386, 265

\bibitem[Strang(1968)]{Strang68}
     Strang, G.\
     1968, SIAM J. Num. Anal., 5, 506 

\bibitem[Strickland \& Blondin(1995)]{SB95}
    Strickland, R. \& Blondin, J.~M.\ 
    1995, \apj, 449, 727

\bibitem[Sutherland et al.(2003)]{SBD03}
    Sutherland, R.~S., Bicknell, G.~V., \& Dopita, M.~A.\ 
    2003, \apj, 591, 238

\bibitem[Thompson(1987)]{Thompson87}
     Thompson, K.~W.\
     1987, JCP, 68, 1 

\bibitem[Thompson(1990)]{Thompson90}
     Thompson, K.~W.\
     1990, JCP, 89, 439 

\bibitem[T{\' o}th \& Draine(1993)]{TD93}
     T{\' o}th, G.~\& Draine, B.~T.\
     1993, \apj, 413, 176

\bibitem[Walder \& Folini(1998)]{WF98} 
     Walder, R.~\& Folini, D.\ 
     1998, \aap, 330, L21 

\bibitem[Wolff et al.(1989)]{Wolff89} 
     Wolff, M.~T., Gardner, J.~H., \& Wood, K.~S.\
     1989, \apj, 346, 833

\bibitem[Wood et al.(1992)]{Wood92}
     Wood, K.~S., Imamura, J.~N., \& Wolff, M.~T.\ 
     1992, \apj, 398, 593 

\bibitem[Wu et al.(1992)]{Wu92}
     Wu, K., Chanmugam, G., \& Shaviv, G.\ 
     1992, \apj, 397, 232 

\bibitem[Wu et al.(1996)]{Wu96}
     Wu, K., Pongracic, H., Chanmugam, G., \& Shaviv, G.\ 
     1996, \pasa, 13, 93 

\bibitem[Wu(2000)]{Wu00}
     Wu, K.\
     2000, \ssr, 93, 611

\bibitem[Yamada \& Nishi(2001)]{YM01}
     Yamada, M.~\& Nishi, R.\ 
     2001, \apj, 547, 99 

\end{thebibliography}
